\documentclass[fleqn,usenatbib]{mnras}

\usepackage{newtxtext,newtxmath}
\usepackage[T1]{fontenc}
\usepackage{ae,aecompl}

\usepackage{graphicx}	% Including figure files
\usepackage{amsmath}	% Advanced maths commands
\usepackage{bm}
\usepackage{float}
\usepackage{dblfloatfix}
\usepackage[normalem]{ulem}
\usepackage{scalerel,tikz}
\usetikzlibrary{svg.path}
\definecolor{orcidlogocol}{HTML}{A6CE39}
\tikzset{orcidlogo/.pic={
 \fill[orcidlogocol] svg{M256,128c0,70.7-57.3,128-128,128C57.3,256,0,198.7,0,128C0,57.3,57.3,0,128,0C198.7,0,256,57.3,256,128z};
 \fill[white] svg{M86.3,186.2H70.9V79.1h15.4v48.4V186.2z}
 svg{M108.9,79.1h41.6c39.6,0,57,28.3,57,53.6c0,27.5-21.5,53.6-56.8,53.6h-41.8V79.1z M124.3,172.4h24.5c34.9,0,42.9-26.5,42.9-39.7c0-21.5-13.7-39.7-43.7-39.7h-23.7V172.4z}
 svg{M88.7,56.8c0,5.5-4.5,10.1-10.1,10.1c-5.6,0-10.1-4.6-10.1-10.1c0-5.6,4.5-10.1,10.1-10.1C84.2,46.7,88.7,51.3,88.7,56.8z};
}}
\newcommand\orcidicon[1]{\href{https://orcid.org/#1}{\mbox{\scalerel*{
\begin{tikzpicture}[yscale=-1,transform shape]
\pic{orcidlogo};
\end{tikzpicture}
}{|}}}}

\title[Baryonic effects on low-mass haloes]{The influence of baryons on low-mass haloes} 

\author[H. Zheng et al.]{%
Haonan Zheng\orcidicon{0000-0002-1665-5138}$^{1,2,3}$\thanks{Email: hnzheng@nao.cas.cn}, 
Sownak Bose\orcidicon{0000-0002-0974-5266}$^{3}$, 
Carlos S. Frenk\orcidicon{0000-0002-2338-716X}$^{3}$, 
Liang Gao\orcidicon{0000-0002-9276-917X}$^{1,2,4,5}$\thanks{Email: lgao@bao.ac.cn},  
Adrian Jenkins\orcidicon{0000-0003-4389-2232}$^{3}$,
\newauthor 
Shihong Liao\orcidicon{0000-0001-7075-6098}$^{1}$, 
Volker Springel\orcidicon{0000-0001-5976-4599}$^{6}$, 
Jie Wang$^{1,2,4}$, 
Simon D. M. White\orcidicon{0000-0002-1061-6154}$^{6}$ 
\vspace*{0.1cm}\\%
$^{1}$Key Laboratory for Computational Astrophysics, National Astronomical Observatories, Chinese Academy of Sciences, Beijing 100101, China\\%
$^{2}$School of Astronomy and Space Science, University of Chinese Academy of Sciences, Beijing 100049, China\\%
$^{3}$Institute for Computational Cosmology, Department of Physics, University of Durham, South Road, Durham, DH1 3LE, UK\\%
$^{4}$Institute for Frontiers in Astronomy and Astrophysics, Beijing Normal University, Beijing 102206, China\\%
$^{5}$School of Physics and Laboratory of Zhongyuan Light, Zhengzhou University, Zhengzhou 450001, China\\%
$^{6}$Max-Planck Institute for Astrophysics, Karl-Schwarzschild Str. 1, D-85748, Garching, Germany
}

\date{Accepted XXX. Received YYY; in original form ZZZ}

\pubyear{2024}

\begin{document}
\label{firstpage}
\pagerange{\pageref{firstpage}--\pageref{lastpage}}
\maketitle

% Abstract of the paper
\begin{abstract}
The Voids-within-Voids-within-Voids (VVV) project used dark-matter-only simulations to study the abundance and structure of dark matter haloes over the full mass range populated in the standard $\Lambda\mathrm{CDM}$ cosmology. Here we explore how baryonic effects modify these results for $z=0$ halo masses in the range $10^4$ to $10^7~\mathrm{M_\odot}$, below the threshold for galaxy formation. Our main study focuses on three simulations from identical initial conditions at $z=127$, one following dark matter only, one including non-radiative gas, and one additionally including the baryonic physics relevant in this halo mass range (cooling and photoheating). In the non-radiative simulation, above $10^{5.5}~\mathrm{M_\odot}$,
halo abundance and internal structure are very similar to the dark-matter-only simulation, and the baryon to dark matter ratio is everywhere close to the cosmic value. At lower mass, this ratio drops and haloes are less concentrated and less massive in the non-radiative case. Test simulations at higher resolution show this to be mainly a resolution effect; the expected drop in baryon content due to residual pressure effects only becomes substantial for $z=0$ haloes below $\sim 10^{2.7}~\mathrm{M_\odot}$. However, gas is heated by reionization at $z=6$ in our ``full physics'' run, and this results in almost complete expulsion of gas from all haloes in our simulated mass range. This suppresses the halo mass function by $\sim 30 \%$, lowers halo concentration, and consequently weakens the dark matter annihilation signal by $\sim 40-60 \%$.  \end{abstract}

% Select between one and six entries from the list of approved keywords.
% Don't make up new ones.

%%%%%%%%%%%%%%%%%%%%%%%%%%%%%%%%%%%%%%%%%%%%%%%%%%

%%%%%%%%%%%%%%%%% BODY OF PAPER %%%%%%%%%%%%%%%%%%

\section{Introduction}
\label{sec:intro}

Measurements of the cosmic microwave background (CMB) have revealed that dark matter is the dominant component ($\sim 84\%$) of cosmic matter \citep{Planck2020}. 
Despite its important role as the principal driver within the $\mathrm{\Lambda CDM}$ framework of the growth both of large-scale structure and of galaxies \citep{Davis1985,White&Frenk1991}, the nature of dark matter is still uncertain, and numerous elementary particle candidates have been proposed (e.g. weakly interacting massive particles a.k.a. ``WIMPs'', sterile neutrinos, axions, see \citealt{Bertone2005} and \citealt{Roszkowski2018} for reviews). 

Various probes involving the present-day properties of small-scale (sub)structures have emerged as possible constraints on the nature of dark matter \citep[e.g. their abundance, concentration, and dark matter annihilation signals, see][for a review]
{Frenk&White2012}. The properties of dark matter particles set the minimum scale for structure in the post-recombination universe, and this in turn determines the minimum mass of the nonlinear structures that  form at later times \citep[e.g.][]{Bode&Ostriker2001, Avila-Reese2003}. A detailed understanding of how this is reflected in the abundance and structure of low-redshift mini-haloes may thus be key in constraining dark matter. 
Until recently, such understanding has come from extrapolating results for much larger haloes, or from approximate structure formation theories \citep[e.g.][]{PS1974,Bond1991,Sheth2001}. Recently, however, the full halo mass range relevant for WIMP dark matter was resolved in the VVV multi-zoom dark matter simulation suite of \citet{Wang2020}. 
This suite uses nine levels of resimulation to span over 20 orders of magnitude in halo mass, resolving present-day halo structure all the way from rich cluster masses ($\sim 10^{15}\ \mathrm{M_\odot}$) down to the Earth mass  limit expected in a typical WIMP model ($\sim 10^{-6}\ \mathrm{M_\odot}$). The VVV simulations thus provide a direct prediction for the abundance and structure of present-day mini-haloes \citep{Zheng2023}. This contrasts with earlier work on low-mass halo formation which concentrated on halo properties at high redshift and did not follow  evolution until the present day \cite[e.g.][]{Diemand2005, Ishiyama2010, Angulo2017}.

These predictions can be improved by taking baryonic physics into consideration, which, as pointed out in many studies, can alter low-redshift halo properties such as abundance (or halo mass) and density profile (or concentration), thus affecting the predicted luminosity in annihilation radiation. 
The effect on halo abundance depends on halo mass and on the baryonic physics considered. 
For example, in simulations with reionization, star formation and supernovae feedback, \citet{Sawala2013} report a $20-30\%$ decrease in the cumulative halo mass function at $10^{9}$ to $10^{11}\ h^{-1} \mathrm{M_\odot}$, and \citet{Grand2021} report a $30\%$ decrease in halo abundance and a factor of two decrease in halo dark matter annihilation luminosity for haloes in the mass range  $10^{7.3}$ to $10^{10.5}\ h^{-1} \mathrm{M_\odot}$
For simulations additionally including black hole feedback, \citet{Schaller2015} report a $20-30\%$ decrease in the halo mass function at $10^8$ to $10^{11} h^{-1}\ \mathrm{M_\odot}$, 
and \citet{Vogelsberger2014} report a similar decrease for $10^{8}$ to $10^{11}\ h^{-1} \mathrm{M_\odot}$. 

We note that the above studies focus on relatively large mass scales ($> 10^{7.3}~\mathrm{M_\odot}$), and cannot be applied to haloes with shallower potential wells and no star formation at all \citep{Rees1986, Thoul&Weinberg1996, Gnedin2000}. 
Recently, \citet{Benitez-Llambay&Frenk2020} presented a detailed model of gas cooling in haloes as a function of redshift and mass, defining a `halo occupation fraction' as the fraction of haloes containing at least some stars as a function of halo mass; they found this halo occupation fraction to drop to zero at $\sim 3 \times 10^{8}\ \mathrm{M_\odot}$. 
This is very convenient for simulations of even lower mass haloes, since only relatively clean baryonic physics need be included, i.e. the cooling and reionization of gas.  

In this study, we carry out high-resolution hydrodynamic simulations, based on the dark-matter-only (DMO) simulations of \citet{Wang2020} in order to investigate the impact of baryons on such low-mass haloes, and how that impact depends on halo mass. 
Our results are useful for making accurate predictions for halo properties relevant to dark matter detection based on methods focusing on mini-haloes, $\lesssim 10^{8} ~ \mathrm{M_\odot}$, such as strong gravitational lensing \citep{Dalal&Kochanek2002, Koopmans2005}, density fluctuations in tidal streams \citep{Ibata2002, Johnston2002,Banik2021}, and $\gamma$-ray emission from dark matter annihilation \citep[e.g.][]{Bergstrom1999, Stoehr2003, Springel2008Nat,Grand2021}. 

The paper is organized as follows. Section~\ref{sec:simulation} describes the simulation details. 
Section~\ref{sec:results} presents our results. Sections~\ref{sec:resolution} and \ref{sec:discussion} discuss 
the impact of thermal pressure, of numerical limitations and of baryonic processes  not included in this paper. Section~\ref{sec:conclusion} presents our conclusions. 

\section{Simulation Details}
\label{sec:simulation} 
To investigate baryonic effects on the properties of mini-haloes we concentrate on three high-resolution zoom simulations with the \textsc{\MakeLowercase{GADGET-4}} code \citep{Springel2020}. We adopt the same cosmological model parameters and generate initial conditions using the same methods as \citet{Wang2020} for the {\small VVV}-project. 
%\textcolor{red}{In order to be able to simulate the population of halos over 20 orders of magnitude in mass, we devised the {\small VVV} strategy whereby we start from a cosmologically representative region and then successively select a nested series xxxxx
%The strategy devised for the {\small VVV} project} 
The {\small VVV} simulations, which model dark matter only,  consist of a single $\Lambda$CDM simulation of a cosmological volume  $737.79\ \mathrm{Mpc}$  on a side (the L0 volume), and a  nested series of zoom simulations within L0 that target smaller and smaller low-density  regions (the L1-L8 volumes). 
In order to be able to simulate the formation of the population of present-day haloes over 20 orders of magnitude in mass with a reasonable computational cost, the strategy adopted in {\small VVV} was to zoom into voids. This, by construction, excludes the most massive virialized objects in the zoom region, reducing the computational expense dramatically. A visual representation of this set-up is shown in Figure~1 of \citet{Wang2020}. Each of these nested zoom simulations contains a core region that can be simulated all the way to redshift zero without being contaminated by heavier particles.  For the current paper it proved convenient to reuse a set of initial conditions created for the {\small VVV} project. We took a region called `L3-pilot' which is located within the L2 core region.  The  high-resolution region of the L3-pilot initial conditions is close both in size and in position to the L3 volume, but it has a particle mass that is 64 times larger than in the {\small VVV}-L3 simulation. 

In Table \ref{tab:table1}, we list the parameters of our three simulations: dark matter only (DM hereafter, identical to the original L3-pilot);  dark matter plus non-radiative primordial gas (NR), and dark matter plus primordial gas that is heated by reionization and able to cool through atomic cooling (RI). We take the DM simulation as the fiducial simulation to compare against.
The initial conditions for the NR simulation is generated from the DM initial conditions using a feature of the  \textsc{\MakeLowercase{GADGET-4}} code that splits each high-resolution dark matter particle into a dark matter particle with less mass and a gas particle. We set the initial gas temperature to 245 K according to the fitting formula provided by \citet{Tseliakhovich&Hirata2010} evaluated at our starting redshift of 127. We adopt a smoothed particle hydrodynamics (SPH) neighbour number, $N_\mathrm{sph}=64$, and the artificial viscosity constant, $\alpha_\mathrm{sph}=1$. 
The RI simulation further includes radiative cooling \citep{Katz1996} and photoheating by a uniform UV/X-ray background from galaxies and quasars \citep{Haardt&Madau1996}. The \textsc{\MakeLowercase{GADGET-4}} code implements cooling and heating assuming ionisation equilibrium at all times.  The UV background is turned on at redshift six triggering prompt reionization of hydrogen at that redshift.
Since atomic cooling has virtually no effect before reionization (the temperature of almost all gas is well below $10^4\ \mathrm{K}$), we adapt a snapshot of the NR simulation at $z=10$ as the initial condition for the RI simulation. 
Baryonic processes such as star formation and stellar feedback are not included, since the minimum halo mass for star formation is $\sim 3 \times 10^{8}\ \mathrm{M_\odot}$ \citep{Benitez-Llambay&Frenk2020}, which is above the mass range present in our simulations. 

\begin{table}
 \caption{Parameters of the different simulations. Column~1: name; column~2: number of high-resolution particles (dark matter and gas); column~3: mass of high-resolution dark matter particles; column~4: mass of gas particles; column~5: softening length of high-resolution particles ($\epsilon_\mathrm{dm} = \epsilon_\mathrm{gas}$); column~6: the distance from the high-resolution centre to the closest higher mass `tidal' particle at $z=0$, representing the size of the high resolution region.}
 \label{tab:table1}
 %\vspace{-0.2cm}
 \begin{tabular}{cccccc}
  \hline
  \shortstack{Name of\\simulations} & \shortstack{\vspace{0.1cm} $N_\mathrm{particles}$} &\shortstack{$m_{\mathrm{dm}}$ \\ $[\mathrm{M_{\odot}}]$} & \shortstack{$m_{\mathrm{gas}}$\\$[\mathrm{M_{\odot}}]$} & \shortstack{$\epsilon_\mathrm{softening}$\\$[\mathrm{cpc}$]} & \shortstack{$r_{\mathrm{res}}$($z=0$)\\ $[\mathrm{ckpc}]$} \\
  \hline

  DM & $4\times 10^7$ & $177.35$ & -- & $29.51$ & $694.45$ \\
  NR & $8\times 10^7$ & $149.48$ &  $27.87$ & $29.51$ & $700.40$ \\ 
  RI & $8\times 10^7$ & $149.48$ &  $27.87$ & $29.51$ & $706.73$ \\ 

  \hline
 \end{tabular}
 
\end{table}

\begin{figure*}
    \centering
	\includegraphics[width=2.0\columnwidth]{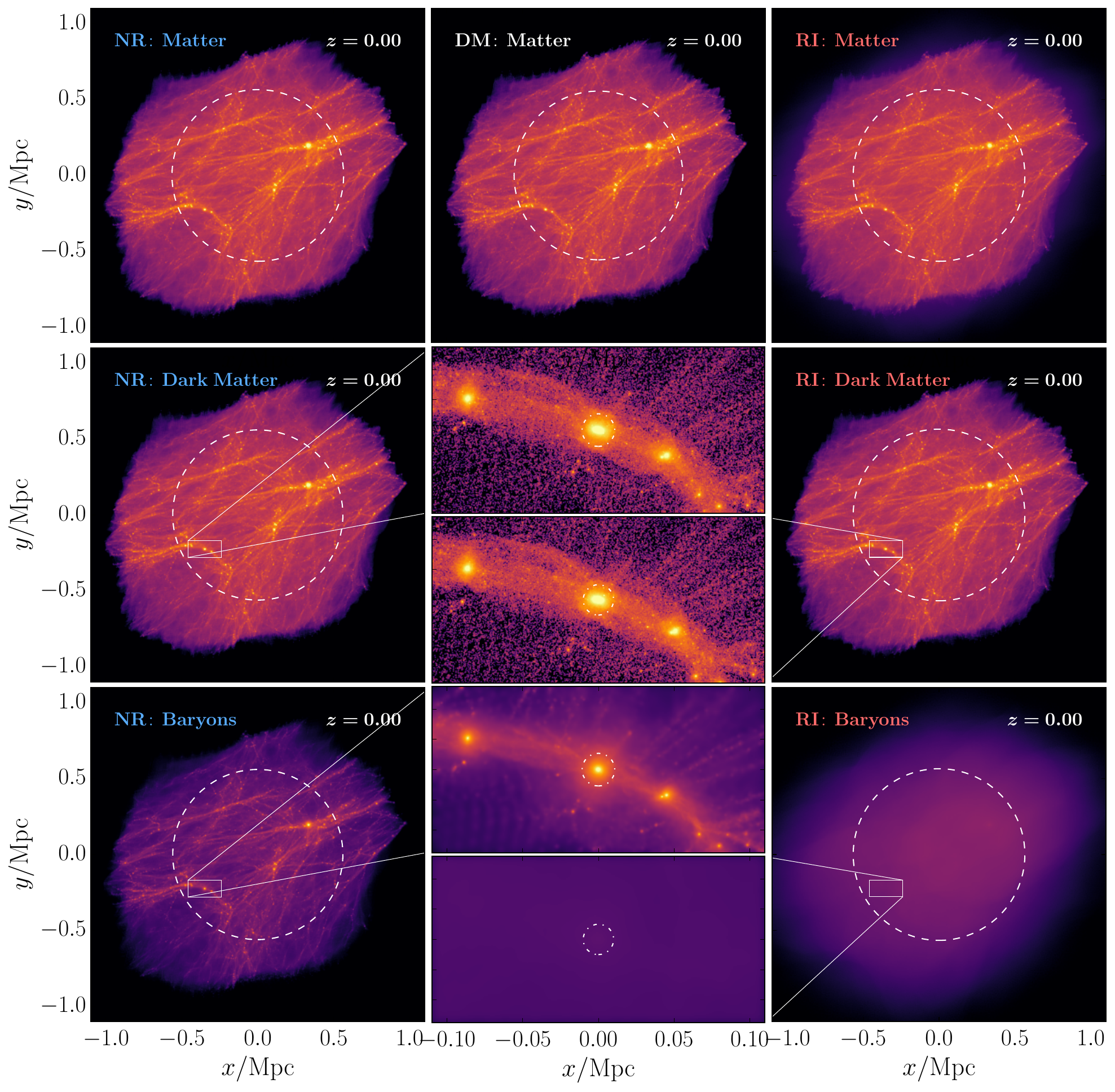}
	\vspace{-0.3cm}
	\caption{{\it Top row:} the projected matter density for the high resolution zoom regions of the three simulations listed in Table~\ref{tab:table1} -- NR, DM and RI.
 {\it Middle row:} left and right columns show the projected dark matter density in the NR and RI simulations. {\it Bottom row:} left and right columns show the projected baryonic density in the NR and NI simulations.  The rectangular images are an enlargement of the region around the largest dark
 matter halo.  The white dashed circles mark the fiducial spherical region we adopt for quantitative analysis as discussed at the end of Section~\ref{sec:simulation}.
	The smaller dot-dashed circles in the rectangular images mark the $r_{200}$ radius of the largest halo in the volume.
 The large-scale structure in the dark matter is very similar among the different simulations. The baryons in the NR simulation mostly follow the dark matter density field, but they are unbound from haloes in the RI simulations. }
	\label{fig:fig1}
\end{figure*}

The cosmological parameters of these simulations are taken from \citet{Planck2014p16}: $\Omega_{\rm{m}} = 0.307$, $\Omega_{\Lambda} = 0.693$, $\Omega_{\rm{b}} = 0.04825$, $h = 0.6777$, $n_\mathrm{s} = 0.9611$ and $\sigma_8=0.8288$. These are the same parameters used in the EAGLE project \citep{Schaye2015}. The same linear matter power spectrum is also used but is extended to higher wavenumbers as explained in  \citet{Wang2020}. \textcolor{black}{We note that our procedures lead to a small inconsistency between our initial conditions and those expected in a Planck cosmology including baryons, since the growth of small-scale linear fluctuations ($k\geq 10^{2.3}~{\rm Mpc}^{-1}$) between recombination and $z=127$ is slightly weaker in a universe containing both dark matter and baryons than in one containing dark matter only, and furthermore the distributions of baryons and dark matter differ slightly on small scales \citep[see, for example, ][]{Delos2023}. This causes us to overestimate slightly the abundance and concentration of haloes at masses below about $10^5~{\rm M_\odot}$, but we prefer our simpler set-up since all differences between the DM run and the two runs with baryons are then due to baryonic effects during the simulated time period.} 

We use \textsc{\MakeLowercase{GADGET-4}} to identify haloes and subhaloes, and to construct halo merger trees. Dark matter haloes are identified using  the ``friends-of-friends'' (\textsc{\MakeLowercase{FOF}}) algorithm \citep{Davis1985}, with a dimensionless linking length of 0.2 and a minimum of 32 dark matter particles. Gas particles are attached to these haloes as the secondary link types. The \textsc{\MakeLowercase{SUBFIND}} algorithm \citep{Springel2001} is then used to identify gravitationally bound subhaloes with at least 20 particles. In this paper, only central haloes are considered. We adopt $M_\mathrm{200}$ to define the mass of haloes, where $M_\mathrm{200}$ is the mass within a sphere centred on the potential minimum with a mean enclosed density of 200 times the mean matter density of our Universe. The merger tree construction follows \citet{Springel2005}, and the merger trees provide the mass accretion history of haloes. 

 For each simulation output we conservatively define a fiducial region as a sphere of radius, $r_{\rm high}$ containing only high-resolution particles and with no larger mass particles close by. Specifically, we take the centre of the sphere to be the centre of mass of all the high-resolution particles and its radius to be $r_{\rm high}=0.8r_{\rm res}$, where $r_{\rm res}$ is the distance from the centre to the closest more massive particle. We use only particles or haloes whose centre lies within this sphere for all our quantitative analysis. In practice, at any given output time the spherical regions in the DM, NR and RI simulations are almost identical allowing accurate like-with-like comparisons to be made.  

\begin{table*}
 \caption{The evolution of rescaled matter density, $\tilde{\rho}=\rho\,\!/\,\!\Omega\,\rho_\mathrm{crit}$, inside a sphere of radius, $r_\mathrm{high}$, in different simulations. Column 1: redshift, $z$; column 2: name; column 3: radius of the  high-resolution region analyzed, $r_\mathrm{high}$; column 4: total matter density, $\tilde{\rho}_{\mathrm{tot}}$, in each region rescaled by the cosmic mean matter density, $\tilde{\rho}_{\mathrm{tot}}=(\rho_{\mathrm{dm}}+\rho_{\mathrm{b}})\,\!/\,\!(\Omega_\mathrm{dm} + \Omega_{\mathrm{b}})\,\rho_\mathrm{crit}$; column 5: rescaled dark matter density, $\tilde{\rho}_\mathrm{dm}=\rho_\mathrm{dm}\,\!/\,\!\Omega_\mathrm{dm}\,\rho_\mathrm{crit}$; column 6: rescaled baryon density, $\tilde{\rho}_\mathrm{b}=\rho_\mathrm{b}\,\!/\,\!\Omega_\mathrm{b}\,\rho_\mathrm{crit}$. }
 \label{tab:table2}
 
  \begin{tabular}{cccccc}
  \hline
  $z$ & Name of simulations & $r_\mathrm{high}$ $[\mathrm{ckpc}]$ & $\tilde{\rho}_\mathrm{tot}$  & $\tilde{\rho}_\mathrm{dm}$ & $\tilde{\rho}_\mathrm{b}$ \\
  \hline

  \     &  DM & 555.56  & 0.07474 & 0.07474 & -- \\
  0  &  NR & 560.32 & 0.07358 & 0.07359 & 0.07355 \\
  \     &  RI & 565.39 & 0.07195 & 0.07309 & 0.06585 \\

  \hline
  
  \     &  DM & 373.72 & 0.27287 & 0.27287 & -- \\
  3.06  &  NR & 375.65 & 0.27018 & 0.27014 & 0.27041 \\
  \     &  RI & 375.66 & 0.27021 & 0.27116 & 0.26520 \\

  \hline
  
  \     &  DM & 317.36 & 0.41056 & 0.41056 & -- \\
  5.72  &  NR & 318.56 & 0.40728 & 0.40726 & 0.40740 \\
  \     &  RI & 318.59 & 0.40719 & 0.40717 & 0.40728 \\

  \hline
  
  \     &  DM & 283.22 & 0.52991 & 0.52991 & -- \\
  9.27  &  NR & 283.83 & 0.52687 & 0.52685 & 0.52695 \\
  \     &  RI & 283.77 & 0.52705 & 0.52704 & 0.52709 \\

  \hline
 \end{tabular}
 
\end{table*}

\begin{figure*}
    \centering
    \vspace{0.15cm}
	\includegraphics[width=2.0\columnwidth]{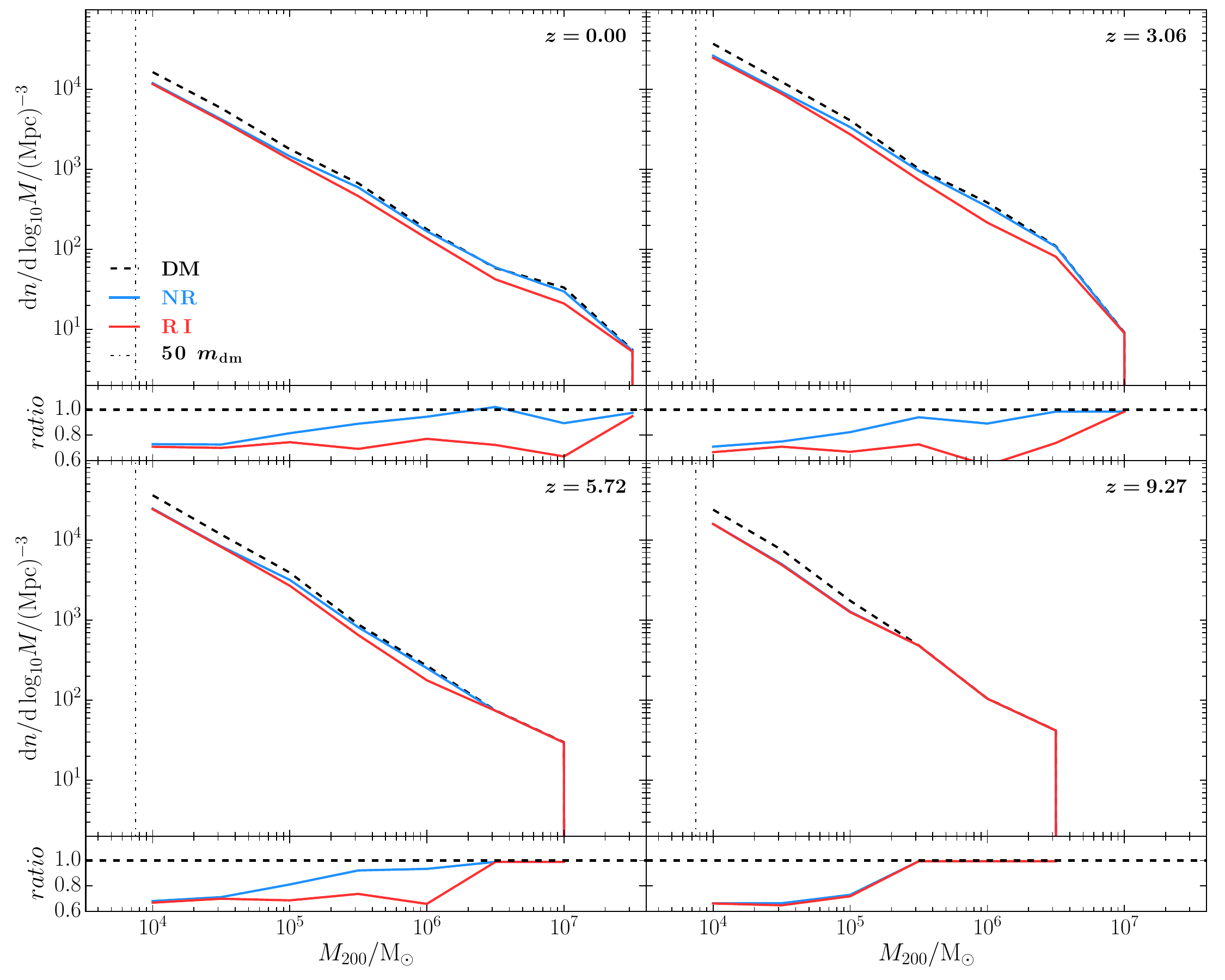}
	\vspace{-0.15cm}
\caption{Halo mass functions in the three simulations at the redshifts shown. At $z=9.27$ the blue solid line is hidden behind the red line. The
vertical dash-dotted lines in the upper 
panel represent the mass of 50 dark matter particles in the hydro runs, indicating the resolution limit above which the halo mass functions are numerically converged within $\sim 20\%$ (as illustrated in Section \ref{sec:resolution}). The smaller panels under each of the mass function plots show the corresponding ratios of the halo mass functions in the NR and RI simulations divided by the reference DM mass function. The halo abundance barely changes at $M_{200} \geq 10^{5.5}~\mathrm{M_\odot}$ in the NR simulation, but it is suppressed by $\sim 30\%$ in the RI simulation since reionization at $z=6$.} 
	\label{fig:fig2}
\end{figure*}

\begin{figure*}
    \centering
	\includegraphics[width=2.0\columnwidth]{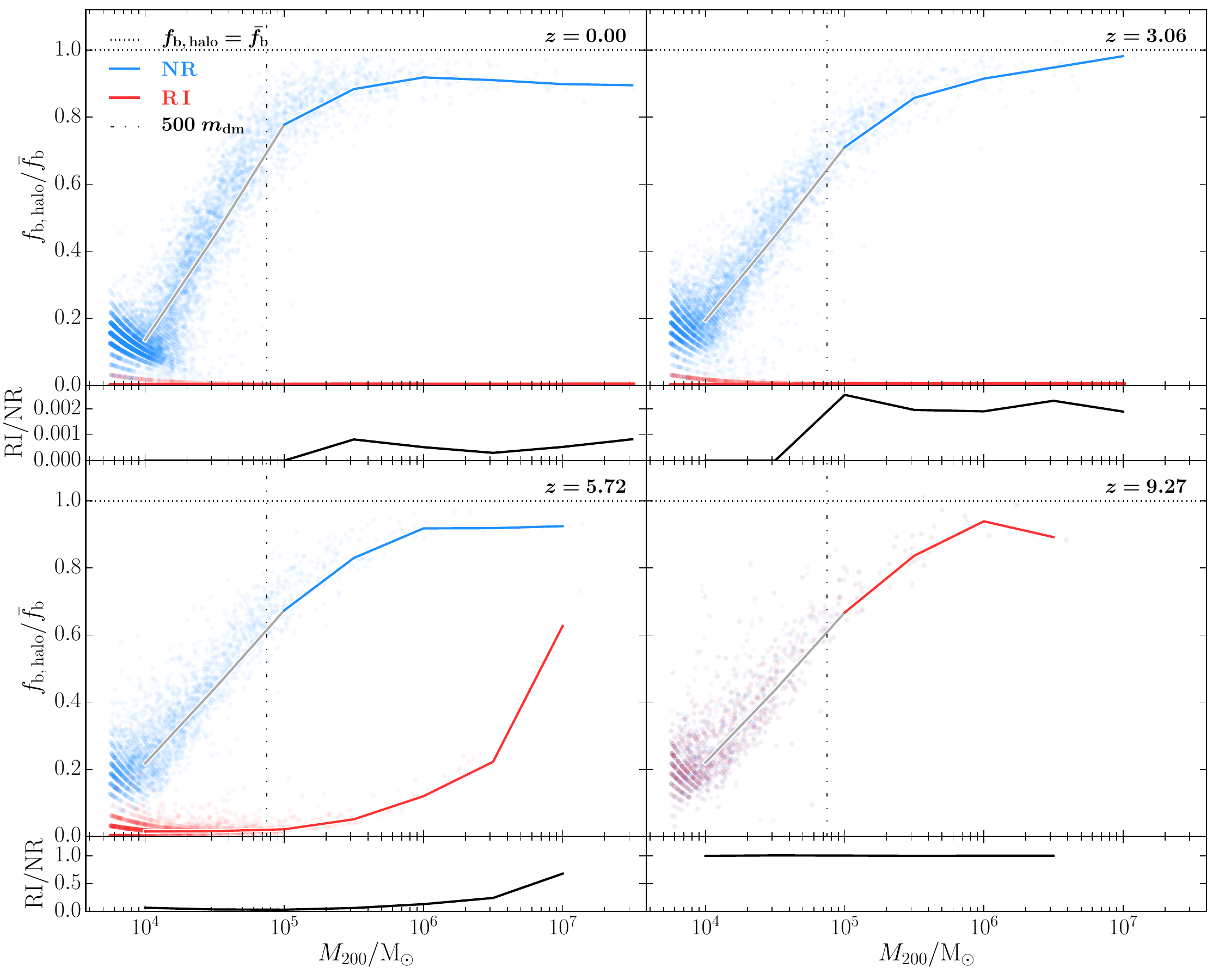}
	\vspace{-0.15cm}
	\caption{The fraction by mass of baryons in haloes within $r_{200}$ as a function of halo mass in the NR and RI simulations at the four redshifts shown. The solid coloured lines show the median fractions and the black horizontal dotted line marks the cosmic mean baryonic fraction, $\bar{f}_\mathrm{b}=\Omega_\mathrm{b}/\Omega_\mathrm{m}$. The vertical dash-dot-dotted lines in the upper panels of each pair represent the mass of 500 dark matter particles in the hydro runs, indicating the resolution limit above which the baryonic fractions in the non-radiative simulations are converged within $\sim 20\%$ (as illustrated in Section \ref{sec:resolution}). To the left of these lines our results are susceptible to numerical issues, so the medians are plotted in grey. The black solid lines in the small panels show the corresponding ratio of the RI/NR baryonic fractions. At redshift $z=9.27$, which is before reionization, the blue NR line is hidden behind the red RI solid line. In the NR simulation, haloes of mass $\geq 10^{5.5}~\mathrm{M_\odot}$ have a baryonic fraction close to the cosmic mean value, while in the RI simulation they lose almost all their baryons after reionization at $z=6$. The faint points in each panel refer to individual haloes. 
 }
	
	\label{fig:fig3}
\end{figure*}

\section{Results}
\label{sec:results} 
\subsection{Large-scale structure}
\label{sec:large-scale}

\begin{figure*}
    \centering
	\includegraphics[width=2.0\columnwidth]{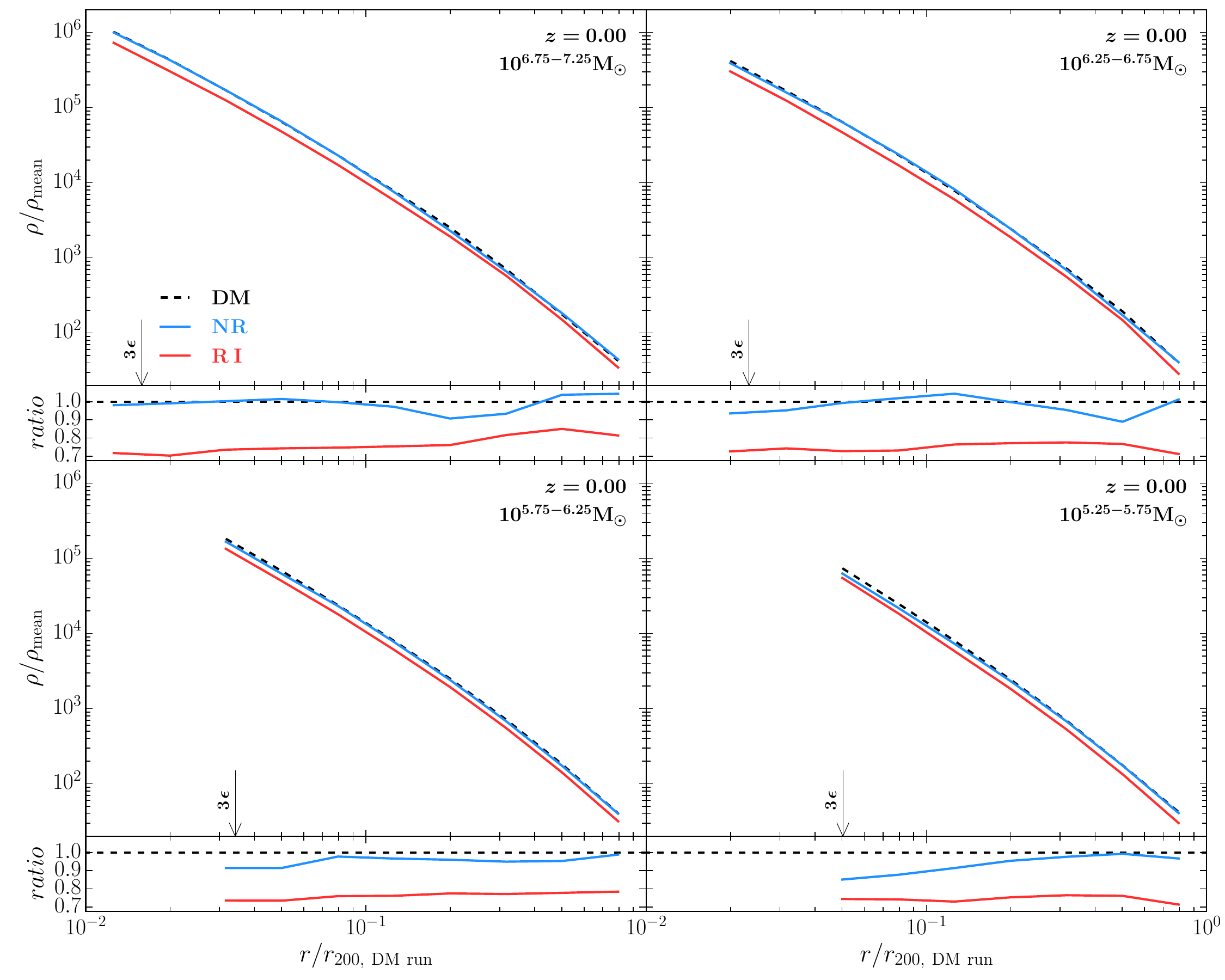}
	\vspace{-0.15cm}
	\caption{Comparison of total matter density profiles of the matched haloes in our three simulations at $z=0$. $\rho_\mathrm{mean}$ denotes the mean matter density of our Universe. Different panels correspond to different mass bins for haloes in the DM simulation. The upper subpanels show density profiles, with the black dashed lines representing the DM simulation, and  the blue and red solid lines the NR and RI simulations, respectively; the arrows mark 3$\epsilon_\mathrm{softening}$. The bottom panels show the ratios NR/DM and RI/DM. 
    For halo masses above $10^{5.5} \mathrm{M_\odot}$ the total matter density profiles are $\sim 15-30\%$ lower for RI than for NR and DM, while below $10^{5} \mathrm{M_\odot}$ both the RI and NR profiles are below the DM profile by a similar amount.     
    }
	
	\label{fig:fig4}
\end{figure*}

We show the $z=0$ projected total matter, dark matter and baryonic matter surface densities in the high resolution regions of our three zoom simulations in Fig. \ref{fig:fig1}. The dark matter distribution on large scales is virtually the same in all three simulations and the shape and size of the high-resolution region is almost unchanged. 
The distribution of gas in the NR simulation closely traces the dark matter structure, while in the RI simulation, where the the gas has been heated by a UV background, the gas is diffuse everywhere and is not concentrated to any visible extent in any of the dark matter haloes within the volume. In the RI simulation some of the gas has left the high resolution region by redshift zero. This is a result of our not including gas in the low-resolution region, so that there is, in effect, a zero-pressure boundary condition on the high-resolution region.

 In Table~\ref{tab:table2} we list the total, dark matter and baryonic densities within our fiducial regions, normalised by the corresponding cosmic mean densities of these components.  Unsurprisingly, given the selection of the underdense L3-pilot region, the total matter density is much lower than the cosmic mean matter density with the ratio decreasing with time. 
 The scaled dark matter and baryon densities, $\tilde{\rho}_\mathrm{gas,\ NR }$, and $\tilde{\rho}_\mathrm{tot,\ DM}$ agree extremely well at all redshifts in the NR simulation. This contrasts with the RI simulation where the
the difference is  $2.2\%$ at  $z=3.06$ and rises to  $11\%$ at $z=0$. The difference is driven by photoheating which raises the temperature of the gas in the RI simulation abruptly at redshift six, leading to the outflow visible at the edges of the gas distribution in the lower right panel of Fig.\ref{fig:fig1}.

\subsection{The halo mass function and the baryonic mass fractions within haloes}

\label{sec:HMF}

The evolution of the halo mass function and of the halo baryonic fractions are shown  in Figs.~\ref{fig:fig2}  and~\ref{fig:fig3} respectively. Both of these quantities are evaluated within our fiducial analysis regions defined at the end of Section~\ref{sec:simulation}. The number of haloes in each mass bin in the $z=0$ DM simulation is given in column 2 of Table~\ref{tab:table3}. 

In the NR simulation, the halo mass function of mini-haloes ($\lesssim 10^5\  \mathrm{M_\odot}$) is suppressed by $\sim$ $30\%$ relative to the dark matter only simulation, 
but the mass function is hardly affected for larger haloes ($\gtrsim 10^{5.5}\ \mathrm{M_\odot}$). The baryonic fraction in haloes of $M_{200} \lesssim 10^5\  \mathrm{M_\odot}$ decreases with halo mass at all redshifts. 
We discuss this trend further in Section~\ref{sec:resolution} where we also determine the number of particles in a halo required for the mass function and the baryonic fractions to be reliably determined. For the NR simulation, the halo mass function converges to within about 20\% for haloes with more than 50 particles (here $\sim 10^4~\mathrm{M_\odot}$) whereas the halo baryonic fractions converge for 500 particles (here $\sim 10^5~\mathrm{M_\odot}$).

In the RI simulation the effect of  reionization ($z_\mathrm{reionization}=6$) is evident. 
At $z=9.27$, before reionization, the halo mass function and halo baryonic fractions overlap with those of the NR simulation.
After reionization, haloes rapidly lose baryons by photo-evaporation, beginning with the smaller haloes which have shallower gravitational potentials. At $z=5.72$ the largest haloes still retain most of their baryons but at lower redshift ($z=3$ and $0$), all of the haloes are nearly gas-free, and the halo mass function is suppressed by $\sim 30\%$ in all mass bins.

Most previous studies (e.g. \citealt{Crain2007, Schaller2015, Qin2017}) comparing haloes in dark matter only and hydrodynamic simulations have focused on much more massive objects
than we consider here, but a few reach sufficiently low mass that their behaviour is dominated by the same effects that are relevant in our mass range. 
\citet{Schaller2015} found the halo mass function over the range $10^8$ to $10^{11.5} h^{-1} \mathrm{M_\odot}$ in the EAGLE simulations \citep{Schaye2015} to be suppressed by $20-30\%$ while \citet{Grand2021} found the abundance of field haloes in the range $10^{7.3}$ to $10^{10.5} h^{-1} \mathrm{M_\odot}$ in the Auriga simulations to be reduced by 30\% relative to DMO versions evolved from the same initial conditions. 

\begin{figure*}
    \centering
	\includegraphics[width=2.0\columnwidth]{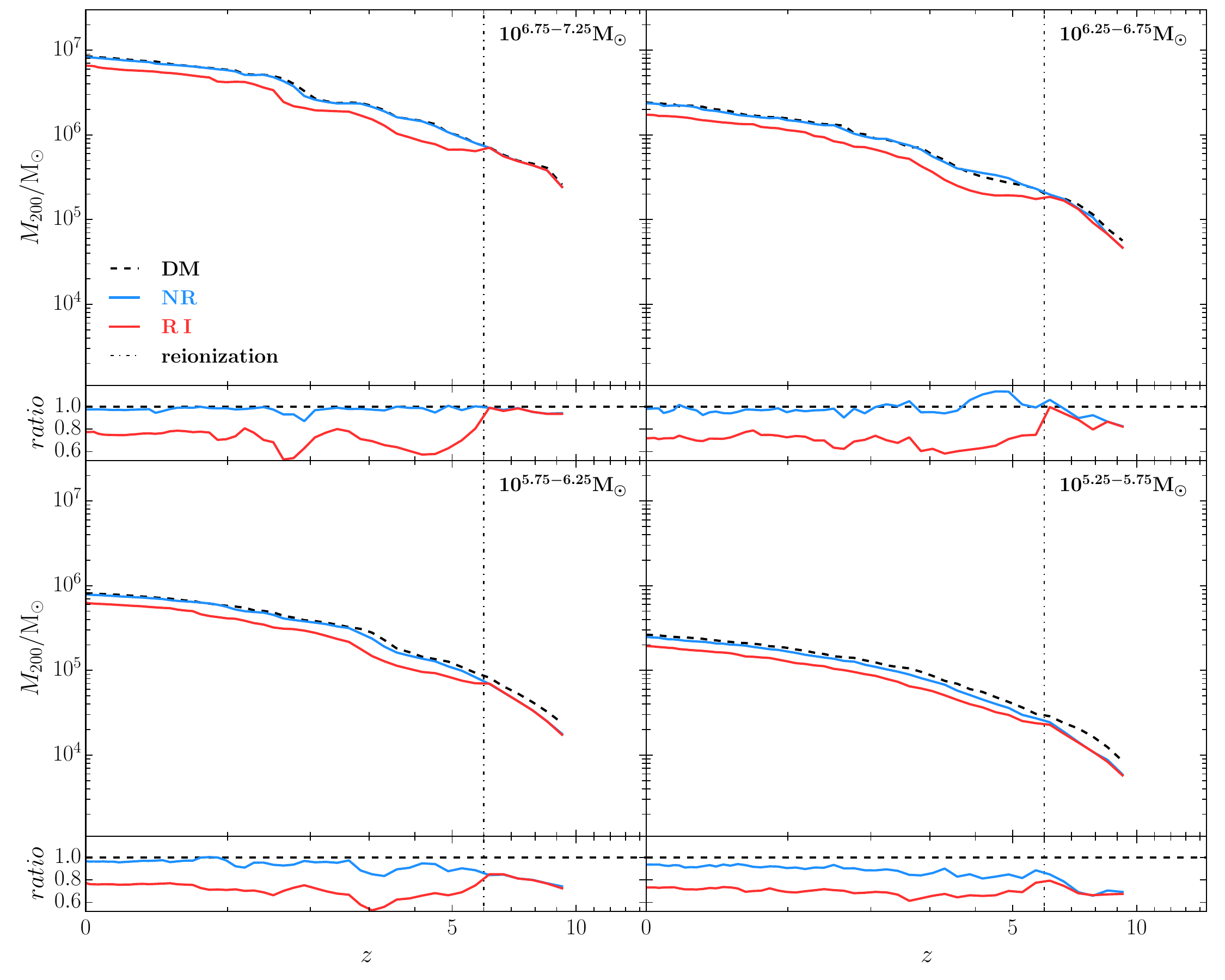}
	\vspace{-0.15cm}
	\caption{Comparison of the mass accretion histories of haloes matched in the three simulations. Different panels correspond to different mass bins in the DM simulation. The upper subpanels show the median mass accretion history, with the black dashed lines representing the DM simulation and the blue and red solid lines the NR and RI simulations respectively; the vertical dash-dotted lines indicate $z_\mathrm{reionization}=6$. The bottom panels show the ratios, NR/DM and RI/DM. Reionization rapidly reduces halo masses by $\sim 30\%$.}
	
	\label{fig:fig5}
\end{figure*}

All simulations that have compared models with non-radiative gas to DMO models from the same initial conditions have found that the abundance and internal structure of haloes are very similar in the two cases. \citet{Crain2007} discussed how post-reionization photoheating affects the baryonic fraction of haloes of mass between $10^{9.5}$ and $10^{13}\ h^{-1} \mathrm{M_\odot}$. 
In their purely non-radiative simulation, the median baryonic fraction within the halo virial radius is $\sim 90\%$ of the cosmic mean value and does not depend either on redshift or on halo mass down to their smallest mass bin ($\sim 10^{9.5}\ h^{-1} \mathrm{M_\odot}$). 
Our results extend this conclusion, showing that above $10^{5.5}\ \mathrm{M_\odot}$, haloes in the NR case retain 90\% of their baryons, while below $10^{4}\ \mathrm{M_\odot}$ the gas fraction is greatly reduced (see Fig. \ref{fig:fig9}). This transition is independent of redshift.
In contrast, in their photoheating simulation, \cite{Crain2007} found that low-mass haloes are unable to keep all their gas, with the baryonic fraction falling to approximately half the cosmic value at $10^{10}\ h^{-1} \mathrm{M_\odot}$, and approaching zero at $10^{9.5}\  h^{-1} \mathrm{M_\odot}$. Similar results at higher redshift were obtained by \citet{Okamoto2008}, with the baryonic fraction in haloes dropping from the cosmic mean value at $\log (M/h^{-1} \mathrm{M_\odot}) \approx 9\ (10)$ to nearly zero for haloes ten times smaller in mass at $z=5.0\ (2.09)$. These results are corroborated by our finding that well after reionization $f_\mathrm{b,\,halo} \approx 0$ for all the haloes in our RI simulation. 

\subsection{Present day density profiles}
\label{sec:profile}

\begin{table}

\caption{Number of haloes (within $r_\mathrm{high}$) in each mass bin at $z=0$ in our three simulations. Column 1: mass bin; column 2: $N_1$, number of haloes in the DM simulation; column 3: $N_2$, number of matched haloes, binned by the mass in the DM simulation; see Section~\ref{sec:profile} for details; column 4: $N_3$, number of matched haloes which are on the main branches of merger trees, binned by the mass in the DM simulation; see Section~\ref{sec:MAH} for details.} 
 \label{tab:table3}
  \vspace{0.2cm}\hspace{1.2cm}\begin{tabular}{cccc}
  \hline
  $\log_{10}(M_{200}/\mathrm{M_\odot})$ & $N_1$ & $N_2$ & $N_3$ \\
  \hline

  7.25-7.75 &    2 &    2 &    2 \\
  6.75-7.25 &   12 &   11 &   10 \\
  6.25-6.75 &   21 &   20 &   20 \\
  5.75-6.25 &   64 &   62 &   60 \\
  5.25-5.75 &  240 &  225 &  223 \\
  4.75-5.25 &  641 &  607 &  602 \\
  4.25-4.75 & 2082 & 1829 & 1801 \\

  \hline
 \end{tabular}
 
\end{table}

In this section we compare the radial density profiles of haloes  in the NR and RI simulations with their counterparts in the DM  simulation. To find counterparts we make use of the common dark matter particle IDs in the three simulations and assume a match if the DM halo and the halo from the other simulation share more than half of the same particles. The third column of
Table~\ref{tab:table3} shows the number of DM haloes in bins of halo mass that have a counterpart in both the NR and RI simulations. Almost all haloes in the mass bins we study are successfully matched. We focus on four mass bins, each of width 0.5 dex, in the mass range ($10^{5.25-7.25}\ \mathrm{M_\odot}$).

In Fig.~\ref{fig:fig4}, the haloes are binned by their mass in the DM simulation, and we compare the bin averaged total matter density profiles of the matched haloes for the three simulations. When comparing the density profiles in the three simulations we must bear in mind that individual profiles are affected near the centre by the gravitational softening and by two-body relaxation \citep{Power2003, Navarro2004}. Recent work by \cite{Zhang2019} suggests that the Power et al. radius \citep{Power2003} can
be too conservative by about a  factor of two. In the plots we mark the radius corresponding to three times the gravitational softening, which for our simulations is close to half the Power et al. radius. 

In the highest mass bin, the density profiles in the NR simulation are virtually unaffected by the inclusion of baryons. There is a trend, however, in all but the highest mass bin for the central densities to be lower than those of the DM counterparts by $5-15\%$. The concentration in the two largest mass bins are barely changed. 

By contrast, the density profiles in the RI simulation are strongly affected in all the mass bins. The total  total matter density at a fixed physical radius is $25-30\%$ lower than in the DM case near the centre and $15-20\%$ lower close to $r_{200}$. 
Fitting the density profiles\footnote{We are only able to fit the halo profiles in two largest mass bins ($10^{6.25-7.25}\mathrm{M_\odot}$) because of resolution limitations: in the smaller haloes, $r_{-2}$, the radius where the logarithmic slope is -2, is smaller than 3$\epsilon_\mathrm{softening}$, which would cause the fits to be unreliable. 
} 
with either an NFW profile \citep{NFW1996} or an Einasto profile \citep{Einasto1965, Navarro2004} with fixed $\alpha=0.16$ following \citet{Wang2020}, we find that the concentration parameter in the RI haloes is also typically reduced by  $15-20\%$.

\subsection{Mass accretion history}
\label{sec:MAH}

We investigate the mass accretion histories of haloes using the same matched samples as the previous subsection -- apart from excluding a small number of haloes that have split off a larger halo in the recent past. This makes only a small difference to the sample sizes which are listed in the fourth column of Table~\ref{tab:table3}. The mass accretion histories, computed by following the main branch of the merger trees (produced by \textsc{gadget-4}) back from the present, are shown in Fig.~{\ref{fig:fig5}}.

For the NR simulation, in the two largest mass bins ($10^{6.25-7.25}\ \mathrm{M_\odot}$),  haloes follow very similar mass accretion histories 
to their counterparts in the DM simulation, showing that just including non-radiative gas  makes little difference.  For the two smaller mass bins ($10^{5.25-6.25}\ \mathrm{M_\odot}$), haloes in the NR simulation are only $\lesssim 5\%$ smaller than their counterparts in the DM simulation at $z=0$, while the difference is much larger, $\sim 30\%$, at earlier redshifts (e.g. $z=6$). 
The agreement between the NR and DM mass accretion histories improves once the halo mass exceeds $10^{5.5}\ \mathrm{M_\odot}$ regardless of the redshift when this occurs,
which suggests that this effect is numerical rather than physical in the non-radiative case. 
This is discussed further in Section~\ref{sec:resolution}. 

For the RI simulation, the history before reionization is almost the same as in the NR simulation, suggesting that cooling processes have no significant impact on haloes of this mass. However, as soon as reionization is triggered, haloes in the RI simulation begin to follow different tracks from their NR counterparts; this reflects the fact that gas is no longer able to accrete onto low-mass haloes, 
and their existing gas content is reduced by photoevaporation. This effect can be seen in the work of \cite{Sawala2013}, who found that haloes of subgalactic scale are 30\% less massive in a RI-type simulation than in the corresponding DMO simulation, a result subsequently corroborated by \cite{Velliscig2014} and \cite{Desmond2017}. 
Our results here extend this conclusion to even smaller haloes (i.e. $10^{4-7}\ \mathrm{M_\odot}$).

\subsection{Annihilation signals}

\begin{figure}
    \centering
	\includegraphics[width=1.0\columnwidth]{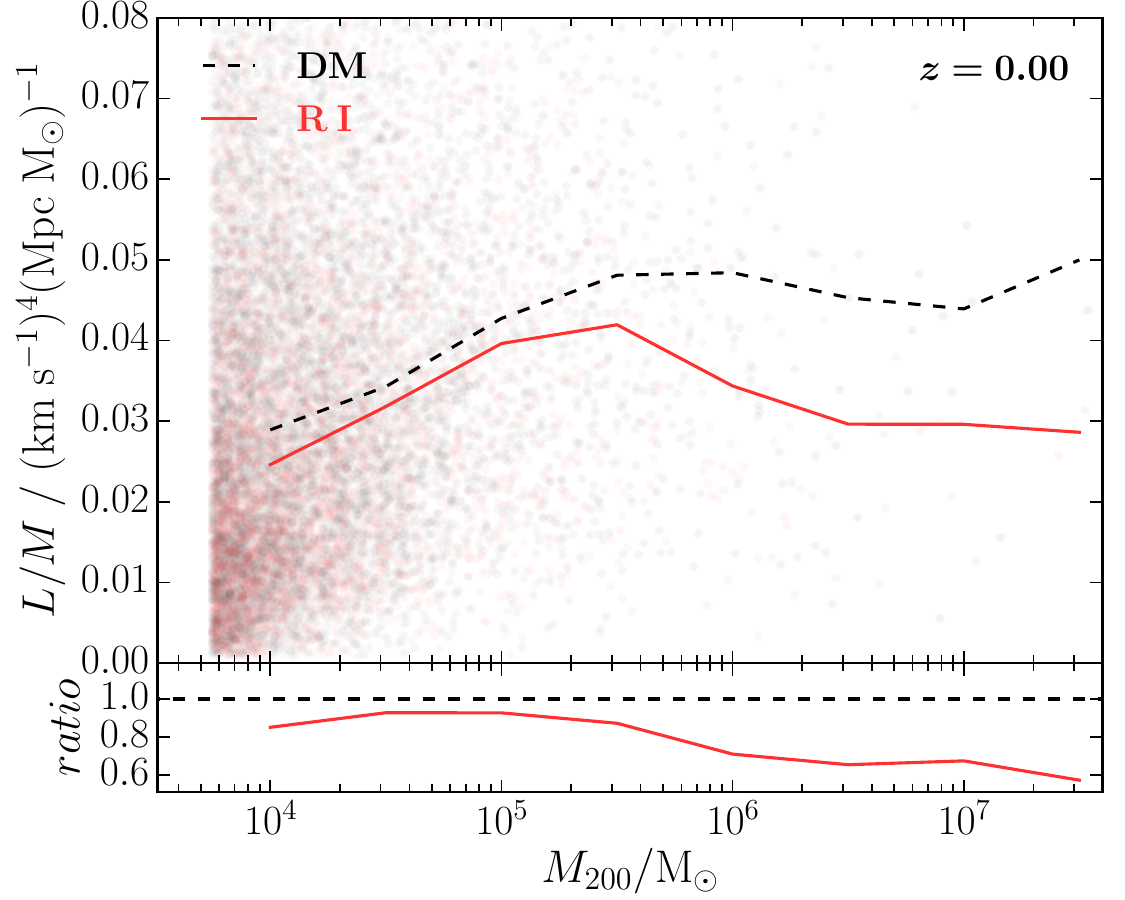}
	\vspace{-0.55cm}
	\caption{Comparison of the annihilation luminosity-to-mass ratio of haloes in the DM (black dashed line) and RI (red solid line) simulations at $z=0$, with the ratio of the means from the top panel shown as a function of halo mass in the bottom panel. We plot mean lines down to $50\, m_\mathrm{dm}$ because such low-mass haloes do not capture significant photoheated gas and so are not affected by possible numerical issues with the hydrodynamics. This $L/M$ ratio is depressed by $\sim 10-40\%$ in the RI simulation. }
	\label{fig:fig6}
\end{figure}

We have found that the presence of baryons reduces both the abundance and the concentration of haloes. Both effects 
lead to a reduction of any annihilation signal coming from the smooth component of haloes and subhaloes.
We follow \citet{Wang2020}, estimating the luminosity per unit mass of a halo from:\footnote{This formula should be corrected by multiplying by $(1-f_\mathrm{b,\,halo})^2$, as baryons do not contribute to dark matter annihilation. However, in the cases we focus on here (i.e. comparing the RI and DM runs at $z=0$), all haloes are almost empty of baryons. }
\begin{equation}
    L/M \propto V^4_\mathrm{max}/(r_\mathrm{max}M_{200}). 
\end{equation}
Here, the halo maximum circular velocity and the radius where this maximum occurs are denoted by $V_\mathrm{max}$ and $r_\mathrm{max}$, respectively.

\begin{figure}
    \centering
    \vspace{0.0cm}
    \begin{minipage}[b]{1.0\columnwidth}
	    \includegraphics[width=1.0\columnwidth]{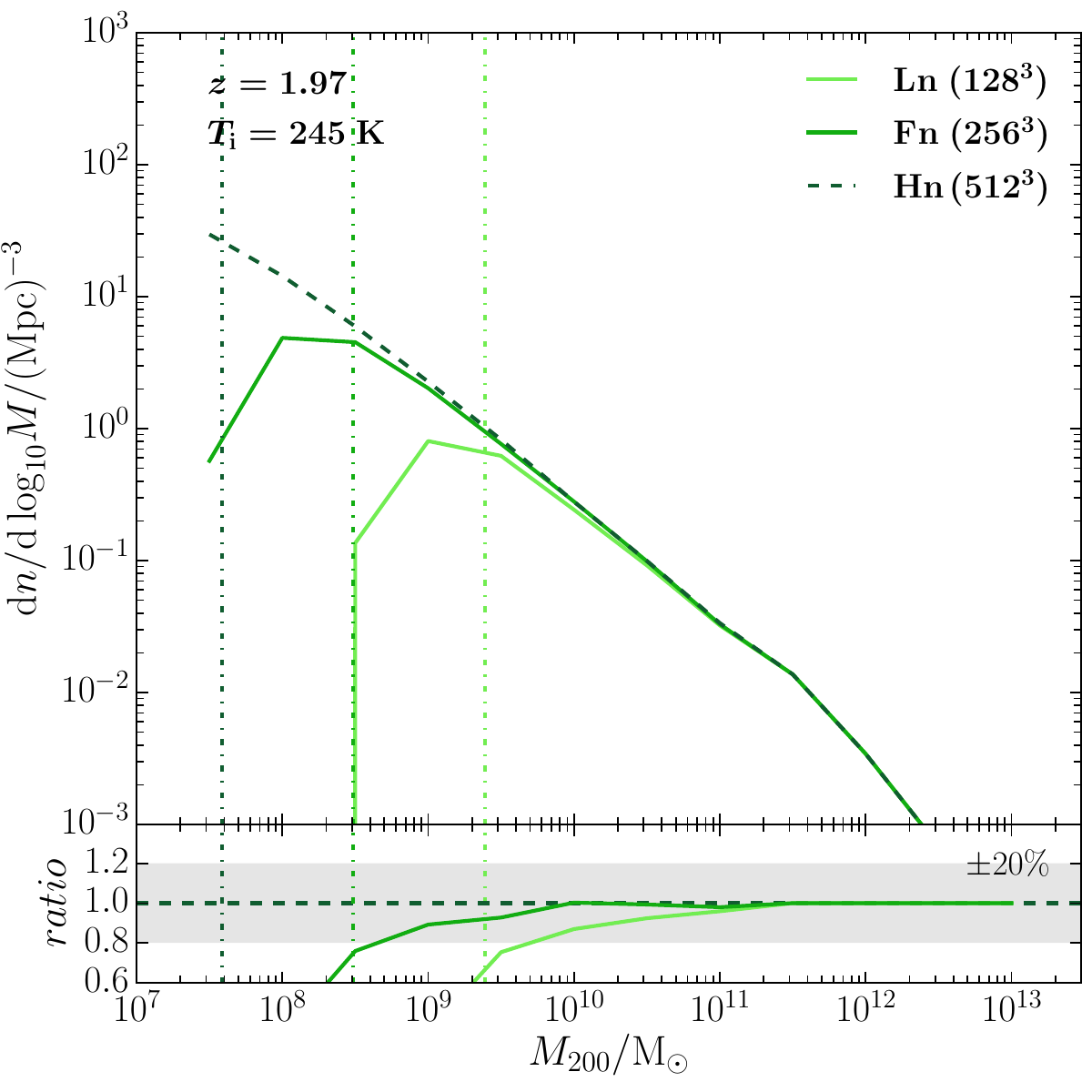}
    \end{minipage}
    \begin{minipage}[b]{1.0\columnwidth}
        \vspace{-0.15cm}
        \includegraphics[width=1.0\columnwidth]{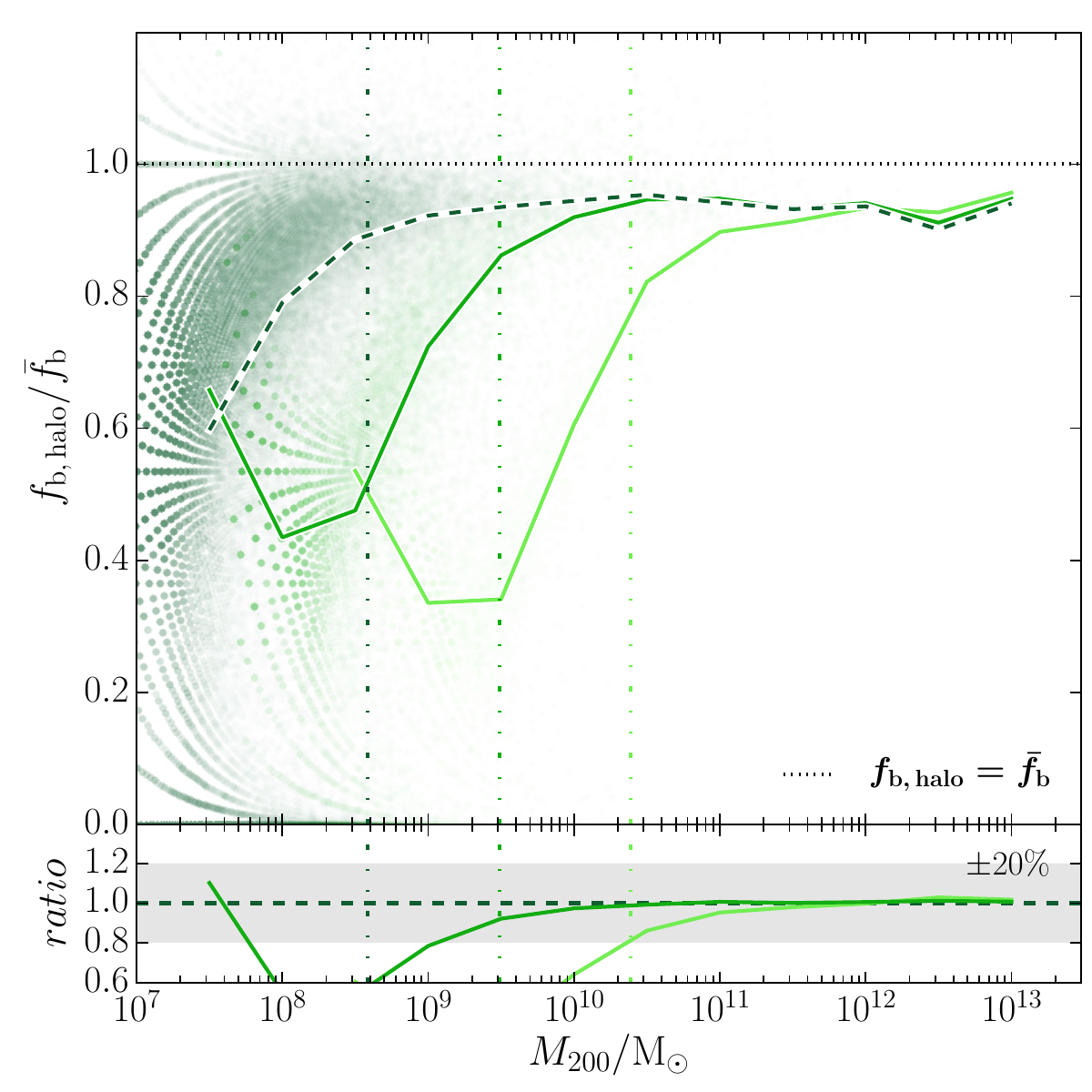}
    \end{minipage}
    \vspace{-0.60cm}
	\caption{Resolution study for the  halo mass function and
          baryonic fraction at $z=2$. We compare results from the Ln
          ($128^3$ dark matter particles, light green), Fn ($256^3$,
          green) and Hn ($512^3$,
          deep green) simulations. These all have the same initial temperature,
          $T_\mathrm{i}=245~\mathrm{K}$, as our fiducial simulations. The subpanels show
          the ratios between the Ln (or Fn) and the Hn simulations. The
          vertical dash-dotted lines in the upper 
          panel represent 50 dark matter particles in each simulation and the
          dash-dot-dotted lines in the bottom panel 500 particles. The
          horizontal dotted line indicates the cosmic baryonic
          fraction, $\bar{f}_\mathrm{b}=\Omega_\mathrm{b}/\Omega_\mathrm{m}$, 
          and the horizontal dashed line + gray shaded band a 
          ratio (relative to the highest resolution simulation) of
          $1\pm 20\%$. The continuous lines in the lower panel indicate median values as a function of halo mass, while the dots in the larger subpanel indicate values for individual haloes with depth of shading representing the relative number of haloes with each value. Note that discreteness effects are visible in this distribution at the lowest halo masses because of the small number of particles involved. The halo mass function is
        converged within $20\%$ in the various 
        simulations  for 
        $M \gtrsim 50 m_\mathrm{dm}$, while the baryonic fraction
        is converged for $\sim 500 m_\mathrm{dm}$.
        }

	\label{fig:fig7}
    \end{figure}

In Fig.~\ref{fig:fig6} we quantify the reduction in $L/M$ for mini-haloes in the RI simulation compared to the DM simulation. We see from the lower subpanel that the signal  decreases by $\sim 40\%$ in the mean for $M_{200} \gtrsim 10^6\ \mathrm{M_\odot}$ (though we note there are only a few haloes in these mass bins), and by $\sim 10\%$ at $M_{200} \lesssim 10^{5.5}\ \mathrm{M_\odot}$. Convolving with the $\sim 30\%$ suppression of the halo mass function in the RI simulation, we estimate that the annihilation signal per unit volume in the RI case is $\sim 40-60\%$ lower than in the DM-only case when averaged over haloes in the mass range $10^4$ to $10^7 \mathrm{M_\odot}$. 
      
\section{Resolution and gas pressure effects in simulations with non-radiative gas}
\label{sec:resolution}

We expect that haloes in our NR simulations that are not resolved with
a sufficient number of SPH particles will have artificially low gas
fractions. There is also a physical effect arising from the choice of
the initial entropy of the non-radiative gas which leads to the gas
being too hot for small haloes to be able to capture or retain it. In
this subsection we make use of additional non-radiative simulations to
separate the numerical and physical effects and to establish their relative
importance for the results presented in Section~\ref{sec:HMF}.

In Appendix~\ref{sec:app} we derive a simple expression for a
characteristic halo mass, $M_{1/2}$, (eq.~\ref{eq:mhalf}) below which we
expect haloes to be largely gas-free because of the effects of gas
pressure. This scale is a function of redshift and of the initial entropy of the non-radiative gas, or equivalently of the temperature of the gas in our initial conditions at $z=127$. 

We make use
of a suite of cosmological simulations within $14.3\ \mathrm{Mpc}$ periodic boxes 
from \citet{Liao2017} to test this characteristic mass scale. These are suitable for our purposes because they include simulations differing only in
the gas temperature in the initial conditions. The parameters of the
simulations are listed in Table~\ref{tab:table4}. The values of the
cosmological parameters are close to, but not identical to those of
the VVV simulations and we do not expect the differences to be
important for our purposes.

\begin{table}
 \caption{Parameters of additional simulations run to test the convergence of halo gas fractions with resolution and choice of initial conditions. Column 1: name of the simulation; column 2: number of dark matter particles; column 3: mass of the dark matter particles; column 4: mass of the gas particless; column 5: softening length; column 6: initial temperature of gas particles at $z_\mathrm{init}=127$. }
 \label{tab:table4}
 %\vspace{0.2cm}
 \begin{tabular}{cccccc}
  \hline
  \shortstack{Name of \\ simulations} & \shortstack{\vspace{0.1cm} $N_\mathrm{dm}$} &\shortstack{$m_{\mathrm{dm}}$ \\ $[\mathrm{M_{\odot}}]$} & \shortstack{$m_{\mathrm{gas}}$\\$[\mathrm{M_{\odot}}]$} & \shortstack{$\epsilon$\\$[\mathrm{kpc}$]} &  \shortstack{$T_\mathrm{i}$\\$[\mathrm{K}$]} \\
  \hline

  Ln & $128^3$ & $4.92\times 10^7$ & $7.56\times 10^6$ & $2.86$ & $245$ \\
  Fn & $256^3$ & $6.15\times 10^6$ & $9.45\times 10^5$ & $2.86$ & $245$ \\ 
  Hn & $512^3$ & $7.69\times 10^5$ & $1.18\times 10^5$ & $2.86$ & $245$ \\ 
  Lf & $128^3$ & $4.92\times 10^7$ & $7.56\times 10^6$ & $2.86$ & $10^7$ \\
  Ff & $256^3$ & $6.15\times 10^6$ & $9.45\times 10^5$ & $2.86$ & $10^7$ \\
  Hf & $512^3$ & $7.69\times 10^5$ & $1.18\times 10^5$ & $2.86$ & $10^7$ \\ 

  \hline
 \end{tabular}
\end{table}

We test numerical convergence in halo properties in the absence of significant effects from the initial gas pressure using the Ln, Fn
and Hn simulations, which have the same initial redshift and temperature (i.e. $z_\mathrm{i}=127$, $T_\mathrm{i}=245\,\mathrm{K}$) as our fiducial simulations but substantially lower resolution.. 
We define convergence
to be agreement within 20\% with a simulation of higher resolution.
Fig.~\ref{fig:fig7} shows that the halo mass function converges at a
mass corresponding to $\sim 50$ dark matter particles. 

\begin{figure}
    \centering
    \vspace{-0.12cm}
    \includegraphics[width=1.0\columnwidth]{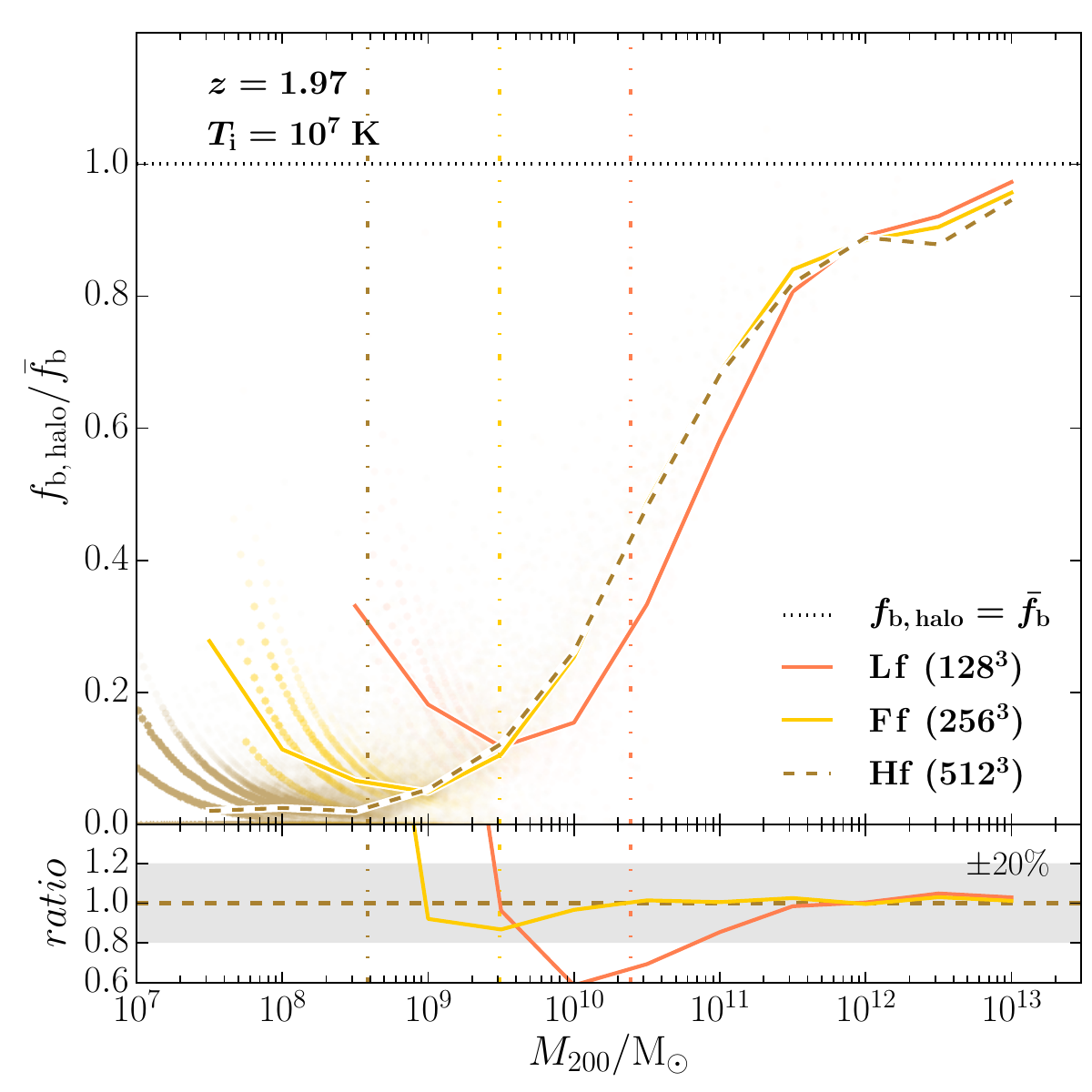}
	\vspace{-0.55cm}
	\caption{As the lower plot of Fig.~\ref{fig:fig7}, but for 
          simulations with a very high initial temperature,
          $T_\mathrm{i}=10^7\ \mathrm{K}$. These simulations are
          Lf ($128^3$ dark matter particles, coral), Ff ($256^3$, gold) and Hf ($512^3$, brown). 
    With the high gas pressure caused by such an extreme
      initial temperature  
      haloes of mass $\lesssim 10^{11} ~\mathrm{M_\odot}$ are unable
      to retain a baryonic fraction close to the cosmic mean value.
    }
	
	\label{fig:fig8}
      \end{figure}

\begin{figure}
    \centering
    \vspace{-0.1cm}
	\includegraphics[width=1.0\columnwidth]{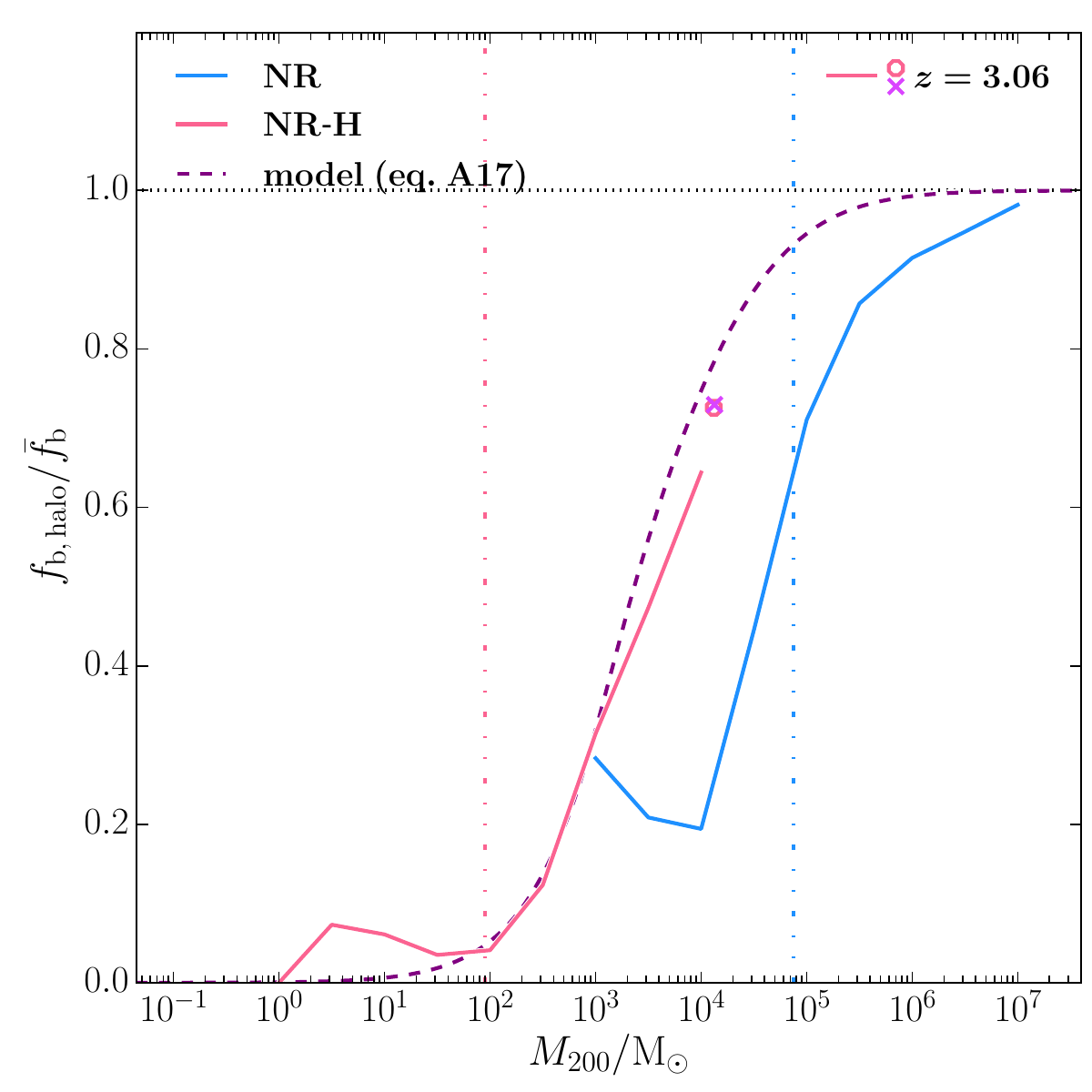}
	\vspace{-0.55cm}
	\caption{Baryonic fraction in the NR and NR-H simulations at $z=3.06$. Blue and pink lines show the median baryonic fraction in the NR and NR-H respectively, while the purple line is the prediction from our model (elaborated in Appendix \ref{sec:app}) assuming halo concentration $c=10$. 
    Thresholds of $500m_\mathrm{dm}$ are marked with vertical dash-dot-dotted lines, and the cosmic baryonic fraction is marked with a horizontal dotted line. The pink and magenta markers show the baryonic fraction of the largest halo in the NR-H simulation and of a further zoomed in simulation of this same halo. The high resolution run shows that haloes with sufficiently small mass $\lesssim 10^{3.4} ~\mathrm{M_\odot}$ (at $z=3.07$) are unable to capture gas even without reionization; they are too small, however, to be resolved in our main NR simulation. }
	
	\label{fig:fig9}
\end{figure}

In all three simulations, the baryonic fraction starts to drop below the cosmic value at radii containing fewer then 500 dark matter particles. Even for the  highest resolution case (Hn) this is at halo masses far above those where there is any appreciable drop in our fiducial NR simulation (see Fig. \ref{fig:fig3}). For all these simulations, the model of Appendix \ref{sec:app} predicts that physical effects due to the finite entropy of the gas will lead to a characteristic mass of
$M_\mathrm{1/2}(z=1.97,\ T_\mathrm{i}=245\ \mathrm{K}) = 1.08 \times
10^3~\mathrm{M_\odot}$, well below the mass scale where the
baryonic fraction drops in any of these simulations, indicating that our adopted initial gas temperature is too low to have any physical effect on the
baryonic fractions of haloes, in particular, also in our fiducial NR simulation. 

In Fig.~\ref{fig:fig8}, we show that non-radiative simulations in which the initial gas temperature is much higher, $T_\mathrm{i} = 10^7~\mathrm{K}$, do show an authentic drop in baryonic fraction.
The baryonic fraction resolved
in the Ff simulation ($256^3$) is closely aligned with that in the Hf
simulation ($512^3$). 
Interpolating values of $M_\mathrm{1/2}$, defined by $f_\mathrm{b,\,halo}(M_{1/2})=\bar{f}_\mathrm{b}/2$,  we find $3.41 \times 10^{10}$ and
$3.54 \times 10^{10}\ \mathrm{M_\odot}$ for the Ff and Hf simulations respectively. a factor of just 1.3 larger than our predicted value,
$M_\mathrm{1/2}(z=1.97,\ T_\mathrm{i}=10^7\ \mathrm{K}) = 2.65 \times
10^{10}\ \mathrm{M_\odot}$. The lowest resolution run, Lf, has $500m_\mathrm{dm}$ comparable to the predicted value of $M_\mathrm{1/2}$ and it clearly overestimates the characteristic mass. Thus, inferring the baryonic fraction reliably in a non-radiative
simulation requires the characteristic mass to be resolved with at
least 500 dark matter particles. This is 
consistent with the conclusion
of \citet{Okamoto2008} that the drop in baryonic
fraction that they found prior to reionization below $10^8\ \mathrm{M_\odot}$ ($\sim 500 m_\mathrm{dm}$ in
their `reference' simulation) was a numerical artefact.

The characteristic mass, $M_{1/2}$, as a function of redshift for our main
simulation is $M_{1/2} \sim 492\ \mathrm{M_\odot}(1+z)^{1.5}$ for haloes with a concentration $c=35$ (see Appendix \ref{sec:app}). 
This is well below the minimum requirement of 500 dark matter
particles per halo to avoid numerical suppression of the halo baryonic
fraction at all redshifts of interest here.  We do expect that the
initial gas entropy will affect the baryon
fractions in very small haloes but simulations of higher numerical resolution are
needed to show this.

We carried out such a simulation, NR-H, making use of higher resolution
initial conditions created originally for the VVV project (the L4-pilot simulation). %described in Section~\ref{sec:simulation}
The mass of the dark matter particles in
the high-resolution region of this resimulation is $0.18\ \mathrm{M_\odot}$, while the mass of
gas particles is $0.034\ \mathrm{M_\odot}$ and the softening
length for all particles is $\sim 4.85\ \mathrm{cpc}$.
Once again we took $T_\mathrm{i} = 245~\mathrm{K}$ at $z_\mathrm{i}=127$. The characteristic mass, $M_{1/2}$, is well resolved
at all redshifts in this case, so we would expect small mass haloes with more than
500 dark matter particles to show a genuine suppression of the baryon
fraction due to the initial gas entropy. This NR-H simulation was expensive and
was stopped at $z=3.06$.

In Fig.~\ref{fig:fig9}, we compare the baryonic fraction in the NR and NR-H simulations with our model prediction
at $z=3.06$. The value
$M_\mathrm{1/2,\, simulation}=3.78\times 10^3\ \mathrm{M_\odot}$ found for NR-H matches the model prediction, 
$M_\mathrm{1/2,\,model}=2.35\times 10^3\ \mathrm{M_\odot}$ well, and, indeed, the model matches the full shape of the simulated suppression as a function of halo mass quite accurately.  
We have checked that similar agreement occurs also at redshifts $5.72$ and $9.27$ and we list the characteristic masses in Table \ref{tab:tableA1}. Finally  
we also resimulated the largest halo in the
NR-H simulation at $z=3.06$ with eight times higher resolution. We show the baryonic fractions found for this one halo in the two simulations as symbols in Fig.~\ref{fig:fig9}.  The two symbols agree very well with
each other and with our theoretical prediction. 

In conclusion, we have determined that at least 500 dark matter
particles per halo are needed in non-radiative simulations to obtain
the baryonic fraction of small haloes to better than 20\%. Provided
this condition is
met, the baryonic fraction of haloes is insensitive to large increases in numerical resolution and
thus appears well converged. We have also shown that the effect of
the initial gas entropy on the baryonic fraction of small haloes can be seen, given sufficient
numerical resolution, and that both the dependence of this suppression on halo mass and the characteristic mass at which it occurs are well reproduced by the simple physical model of Appendix \ref{sec:app}.
It is notable that the halo mass function is
less sensitive to resolution; we find that a minimum of just
50 particles is sufficient to determine it in non-radiative
simulations to an accuracy of 20\%.

\section{Discussion}
\label{sec:discussion}
The NR and RI simulations model the baryons in a very simple way. In this section we consider some of the limitations of our
simulations and how the effects of additional physical processes that have not been modelled might affect our results.

\subsection{Applicability of our results to haloes of lower masses}

We have focused on mini-haloes with mass between $10^{4-7}\ \mathrm{M_\odot}$ but we expect in the cold dark matter cosmogony that there will be haloes of much smaller mass down to the (unknown) cut-off in the matter power spectrum. From our model predictions in Appendix \ref{sec:app} and the results of the highest resolution simulations shown in Fig.~\ref{fig:fig9}, we find that the gas fractions of haloes with masses below $\sim 10^{2.7}\ \mathrm{M_\odot}$ (at $z=0$) or $\sim 10^{3.8}\ \mathrm{M_\odot}$ (at $z=9.27$) are very low in non-radiative simulations simply because the gas entropy is too high for such small haloes to have accreted any gas. We expect that this would lead to an even stronger suppression of the low-redshift halo mass function and of low-redshift halo concentrations relative to a dark matter-only simulation than we saw for haloes in the range of $10^{4-7}\ \mathrm{M_\odot}$ in the RI simulation, where reionization drives gas out of all haloes after redshift 6.

\subsection{Effect of early relative streaming motions between baryons and dark matter}

A streaming velocity, $v_\mathrm{bc}$, between dark matter and baryons is generated at early times because the baryons are coupled strongly to radiation much longer than the dark matter.  The typical root-mean-square value of $v_\mathrm{bc}$ at the recombination time is $\sigma_\mathrm{vbc}\sim 28\ \mathrm{km\ s^{-1}}$, and then decays as $1+z$ \citep{Tseliakhovich&Hirata2010}. 
This streaming velocity is thought  to have a major impact on the formation of the first baryonic structures: \citet{Tseliakhovich&Hirata2010} show that it would suppress the halo mass function for $10^{4-8}\ \mathrm{M}_{\odot}$ haloes at $z=40$ by $\sim 50-70\%$. A similar conclusion may be found in \citet{McQuinn&OLeary2012} at $z\sim 15-40$. According to \citet{Greif2011}, \citet{Fialkov2014} and \citet{Schauer2019}, the streaming velocity may hinder the formation of the first stars. 

Although the effects of the streaming velocity are important at high
redshift, they become less so at lower redshift. For example,
\citet{Naoz2012} show that the halo abundance at
$10^{6-7}\ \mathrm{M_\odot}$ at $z=11$ is almost unaffected for
$v_\mathrm{bc}=\sigma_\mathrm{vbc}$; to make a
difference of $\sim 30\%$ at the same redshift, an extreme value of
$v_\mathrm{bc}$ as high as $3.4\sigma_\mathrm{vbc}$ would be required
for an appreciable effect, which would be a rare case according to \citet{Ahn2016} and \citet{Ahn&Smith2018}.
Thus, while
streaming might induce a sizeable
suppression of halo formation and baryonic fractions at $z \gtrsim 20$, the majority of haloes in our void region form much
later than this and we expect only minor changes from our results at
$z<9.3$.

\subsection{Molecular hydrogen cooling}

Another issue that needs addressing is the possibility of Pop III star
formation before reionization. These first stars could produce 
additional feedback that could expel baryons from haloes at early
times. As we focus on an extremely underdense region, halo growth is
suppressed (e.g. in our simulations, the mass of the largest halo only
exceeds $10^6\ \mathrm{M_\odot}$ at $z\sim12$), leading to delayed
Pop III star formation. Most previous studies \citep[e.g.][]{Yoshida2003,
  Reed2005, Gao2007b, Wise2008} have focused on regions of average or
above average density. For example, \citet{Regan2022} performed zoom
simulations of three clusters and one region of cosmic mean density,
and found Pop III star formation before $z=22$ in all three clusters,
while the cosmic mean density region showed a slower halo mass
accretion history and no sign of star formation when the simulation
ended at $z=20.85$.

In their hydrodynamical simulations, \citet{Yoshida2003} found a
critical halo mass, $\sim 5 \times 10^5\ h^{-1} \mathrm{M_\odot}$, for
Pop III star formation to occur at $z>16$ (mostly at the intersections
of filaments). \citet{Gao2007b, Gao2010} suggested a threshold virial temperature, $T_\mathrm{vir} $
$ \sim 1000 ~ \mathrm{K}$, for molecular hydrogen
production to be boosted and cooling to become efficient, allowing the first
stars to form with a redshift delay, $\Delta z \sim 3-4$, in haloes
whose mass increases from $2.21 \times 10^5 ~ \mathrm{M_\odot}$ to
$2.55 \times 10^6 ~ \mathrm{M_\odot}$ between $z \sim 50 - 10$. By
comparison, the largest haloes in our simulation have  masses of 
$M_{200} = 6.01 \times 10^4 ~ \mathrm{M_\odot}$, $4.97 \times 10^5 ~
\mathrm{M_\odot}$ and $3.37 \times 10^6 ~
\mathrm{M_\odot}$,  corresponding to virial temperatures of 
$T_\mathrm{vir} = 326\ \mathrm{K}$, $961\ \mathrm{K}$ and $2531\ \mathrm{K}$ at
$z = 19.90, 14.07$ and 10.08 respectively. This suggests that first star
formation could possibly occur in our simulation at $z \sim 10-14$. 

However, the possibility of  background Lyman-Werner  (LW) radiation released by an earlier
generation of Pop III stars complicates the situation considerably. 
\citet{Yoshida2003} found that even a low value of LW
radiation of $J_{21}=0.01$ (where $J_{21}$ denotes the intensity of
the radiation in units of 
$10^{-21}\ \mathrm{ergs\ s^{-1}\ cm^{-2}\ Hz^{-1}}$) could increase
the critical virial temperature from $1800\ \mathrm{K}$ to
$2800\ \mathrm{K}$ due to the dissociation of molecular hydrogen
by the background photons; this could, however, be largely 
compensated by self-shielding. They
also found that a value of $J_{21}=0.1$ would almost
entirely prevent gas from cooling and collapsing. \citet{Reed2005} 
found a similar result for $J_{21}=0.08$ and argued that shielding of LW
radiation is unimportant in this case. (The transmission
factor is reduced to 0.1 only over a large distance, 
$1\ \mathrm{cMpc}$; by comparison, the radius of our high resolution
region at $z=12$ is $\sim 0.335\ \mathrm{cMpc}$).

In our case, the first stars are very likely to form outside our
underdense high-resolution region before the haloes in our simulations
gain enough mass to form stars. 
First star formation was long considered highly uncertain due to
the many competing effects\footnote{For example: i)~LW radiation;
ii)~X-ray and UV radiation that can increase the free electron
density and so catalyze molecular hydrogen production; iii)~the explosion
of the first SNe, sending blast waves into the surrounding
gas, removing much of the gas from surrounding haloes and  preventing further cooling; iv)~metal enrichment
which enhances cooling; v)~$\mathrm{H^{-}}$ photodetachment, which
suppresses the production of molecular hydrogen. } \citep{Bromm&Larson2004, Reed2005}, 
but some recent studies, integrating these effects within more sophisticated subgrid models, suggest a clearer conclusion: the earliest-forming first stars inhibit further star formation elsewhere. 
For example, \citet{Nebrin2023}, 
using the LW background provided by the model of \citet{Incatasciato2023}, 
% Fig. 9
found that the halo mass threshold to form the first star is increased by $\sim 0.5-1$ dex between $z=15-10$ due to this radiation; 
\citet{Hegde2023} 
% their Fig. 12
suggested a similar mass threshold of $\sim 5 \times 10^6 ~(10^6) ~\mathrm{M_\odot}$ at $z=10~(15)$ based on a semi-analytic method, particularly when using the same model for the gas central density as \citet{Nebrin2023} (their eq. 7). 
%Thus, whether or not Pop III stars could form in our simulation may depend crucially on the influence of regions outside our simulation. 
These results indicate that most of the slowly-growing haloes in our study miss the time window for forming first stars, further justifying our decision to neglect star formation and related subgrid physics in our simulations.

\section{Conclusion}
\label{sec:conclusion}

In this paper we study the properties of mini-haloes
($10^{4-7}\ \mathrm{M_\odot}$) using a suite of very high resolution hydrodynamic
simulations 
($m_\mathrm{dm} \sim 150\ \mathrm{M_\odot}$, $m_\mathrm{gas} \sim 28\ \mathrm{M_\odot}$). This work builds upon the
methods and results of the VVV multi-zoom simulation project
\citep{Wang2020} that modelled the formation of haloes of all masses
in a dark matter only universe up to the
present day. 
Our simulations consist of a ``DM simulation'' (only dark matter), an
``NR simulation'' (non-radiative gas), and an ``RI simulation'' where
radiative cooling and photoheating of gas by a UV background is turned
on at $z=6$ leading to prompt reionization. We make use of initial
conditions created originally for the VVV project and
model the formation of mini-haloes in a low density region
approximately 1 Mpc across today.

On the scale of the region as a whole, the large-scale structure at
the present day is very similar in all three simulations. The gas in
the NR simulation traces the dark matter closely at all times.
However, after reionization the gas in the RI simulation is heated to
the point that it becomes diffuse and smooth on the scale of even the
largest haloes in the region.

The halo mass function and baryonic fraction are almost identical in
the NR and RI simulations before reionization, demonstrating that
cooling has only a minor effect on haloes in this mass range. After
$z_\mathrm{reionization}=6$, gas flows out of all haloes in the RI
simulation leading to a suppression of the halo mass function by
$\sim 30\%$ at $z=0$.

In the NR simulation we see a drop in the baryonic fraction at
$ M_{200} \sim 10^{5.5}\ \mathrm{M_\odot}$, but we conclude that this is
mostly a numerical artefact. In Section~\ref{sec:resolution} we show that the
baryonic fraction in non-radiative simulations is suppressed by
numerical effects in haloes with less than 500 dark matter particles.
We also show that non-radiative simulations with sufficient resolution
are able to model the physical suppression of the baryonic fraction due to the initial gas entropy. In Appendix~\ref{sec:app} we set up a simple analytic model for this suppression 
as a function of halo mass and redshift and show that it reproduces quite well the evolution found in
simulations with sufficient numerical resolution. Thus, small haloes ($M_{200} \lesssim 10^{2.7} ~ \mathrm{M_\odot}$ at $z=0$; $M_{200} \lesssim 10^{3.8} ~ \mathrm{M_\odot}$ at $z=9.27$) will never be able to capture gas, even in non-radiative runs without reionization. In consequence, the streaming velocity between baryons and dark matter at early times will have little effect on such small haloes. 

We identify corresponding haloes in our three different simulations in order to
study baryon effects on individual haloes, comparing results from
the NR and RI simulations with those from the reference DM simulation.
The density profiles and mass accretion histories
of haloes more massive than $10^{6.25}\ \mathrm{M_\odot}$ are very similar in the NR and
DM simulations, while for smaller haloes,
$10^{5.25-6.25}\ \mathrm{M_\odot}$, the density profile in the
central regions, 0.1$r_{200}$, is $\sim 10\%$ lower in the NR case; in this mass range, halo masses
at high redshift are suppressed by up to $30\%$ in the NR case. In the RI simulation,
the loss of gas from haloes gives rise to shallower density profiles
and a reduction of $\sim 30\%$ in mass at low redshift over the entire mass range, showing that reionization reduces halo potential wells.

The RI simulation allows an improved prediction for dark matter
annihilation signals from the smooth dark matter component of haloes.
We find that the lowered abundance and concentration combine to reduce the annihilation rate per volume for mini-haloes in the mass range
$10^{4-7}\ \mathrm{M_\odot}$ by $40-60\%$.

\section*{Acknowledgements}
We thank the anonymous referee for a very constructive and useful report which helped to improve our manuscript. 
We acknowledge support from the National Natural Science
Foundation of China (Grant No. 11988101, 11903043, 12073002,
11721303) and the K. C. Wong Education Foundation. 
HZ acknowledges support from the China Scholarships 
Council (No. 202104910325). SB is supported by the UK
Research and Innovation (UKRI) Future Leaders Fellowship [grant number
MR/V023381/1]. CSF acknowledges support by the European Research
Council (ERC) through Advanced Investigator grant, DMIDAS (GA
786910). ARJ and CSF acknowledge support from UKRI grant ST/X001075/1. JW acknowledges the support of the
research grants from the Ministry of Science and Technology
of the People’s Republic of China (No. 2022YFA1602901), 
the China Manned Space Project (No. CMS-CSST-2021-B02), 
and the CAS Project for Young Scientists in Basic Research (Grant No. YSBR-062). 
This work used the DiRAC@Durham facility managed by the
Institute for Computational Cosmology on behalf of the STFC DiRAC HPC
Facility (www.dirac.ac.uk). The equipment was funded by BEIS capital
funding via STFC capital grants ST/K00042X/1, ST/P002293/1,
ST/R002371/1 and ST/S002502/1, Durham University and STFC operations
grant ST/R000832/1. DiRAC is part of the UK National e-Infrastructure.

\section*{Data availability}
The data presented in this article will be shared upon reasonable
request to the corresponding author.

%%%%%%%%%%%%%  %%%%%%%%%%%%%%%%%%%%%%%%%%%%%%%%%%%%%
\bibliographystyle{mnras}
\bibliography{main} %

\appendix

\section{The effect of initial entropy on halo gas density profiles in non-radiative simulations}
\label{sec:app}

Let us consider the entropic function $S$, defined for a non-relativistic gas with pressure $p$ and density $\rho_\mathrm{g}$ by $S\equiv p/\rho_\mathrm{g}^{5/3}$. In the absence of significant heating or cooling, gas evolves adiabatically and $S$ remains constant except at shocks, where it always increases. Thus in non-radiative simulations of the kind studied in this paper, S must everywhere be at least as large as the value, $S_{\mathrm i}$, defined by the initial density and temperature of the gas.

On the other hand, well-resolved non-radiative cosmological simulations in which $S_{\mathrm i}$ is negligibly small produce haloes in which the baryon fraction is close to the cosmological value and the gas density profile is very similar to that of the dark matter, hence to that found in dark matter only simulations. Thus, halo accretion shocks have just the strength needed to produce the profile, $S(r)$, which corresponds to $\rho_\mathrm{g}(r)\propto\rho(r)$ in hydrostatic equilibrium, where $\rho(r)$ is the total mass density. \cite{Taylor&Navarro2001} made the remarkable discovery that for haloes with an NFW total density profile, these conditions require $S(r)$ to be very close to a power law.

In our own non-radiative simulations (NR, NR-H and the simulations of Table~\ref{tab:table4}) we find that
the gas and dark matter densities do track each other in high-mass haloes and at large radii, resulting in near power-law behaviour for $S$, but that $S\approx S_{\mathrm i}$ at small radii where $S<S_{\mathrm i}$ is predicted by inward extrapolation of the large-radius behaviour. This motivates a simple analytic model where haloes are taken to have NFW mass profiles, the gas and dark matter densities are assumed parallel at large radii where this implies $S>S_{\mathrm i}$, and the gas is adiabatic with $S=S_{\mathrm i}$ at smaller radii.

For a spherical system in hydrostatic equilibrium, the potential $\phi(r)$, the gas density $\rho_\mathrm{g}(r)$ and the gas pressure $p(r)$ satisfy
\begin{equation}
\label{eq:eql}
    \frac{\mathrm{d}p(r)}{\mathrm{d}r} = -\rho_\mathrm{g}(r)\frac{\mathrm{d}\phi(r)}{\mathrm{d}r}. 
\end{equation}
Assuming \(\displaystyle \rho_\mathrm{g}(r)=\bar{f}_\mathrm{b}\,\rho(r) \), where $\rho(r)$ is the total mass density profile, and $\bar{f}_\mathrm{b} \equiv \Omega_\mathrm{b}/\Omega_\mathrm{m}$ is a constant, the pressure profile can be obtained by integration, 
\begin{equation}
    p(r) = \bar{f}_\mathrm{b}\int^{\infty}_{r} \rho(r')~\frac{\mathrm{d}\phi(r')}{\mathrm{d}r'} \mathrm{d}r'.
\end{equation}
If we now take $\rho(r)$ and its associated $\phi(r)$ to have NFW form, we have 
\begin{equation}
    \rho(r)=\rho_\mathrm{s}\tilde{r}^{-1}(1+\tilde{r})^{-2},
\end{equation}
the enclosed mass within radius $r$ is
\begin{equation}
    M(r)=4\pi\rho_\mathrm{s}r_\mathrm{s}^3\left[ \mathrm{ln}(1+\tilde{r}) - \frac{\tilde{r}}{1+\tilde{r}}\right],
\end{equation}
and the potential is
\begin{equation}
\label{eq:potential}
    \phi(r) = -4\pi G \rho_\mathrm{s} r^2_\mathrm{s} \tilde{r}^{-1} \ln(1 + \tilde{r}).
\end{equation}
Here $r_\mathrm{s}$ and $\rho_\mathrm{s}$ are the scale radius and the characteristic density of the NFW profile, and $\tilde{r}\equiv r/r_\mathrm{s}$ is the nondimensionalized radius. 

Thus, the gas density profile is 
\begin{equation}
\label{eq:rhogro}
    \rho_\mathrm{g}(r)=\bar{f}_\mathrm{b} \cdot \rho_\mathrm{s} \tilde{r}^{-1}(1+\tilde{r})^{-2},
\end{equation}
the pressure profile is
\begin{equation}
    p(r)=\bar{f}_\mathrm{b} \cdot 4 \pi G \rho^2_\mathrm{s} r^2_\mathrm{s}\cdot K(\tilde{r}),
\end{equation}
and the entropic function profile is
\begin{equation}
\label{eq:sr}
    S(r)\equiv p(r) \rho_\mathrm{g}(r)^{-5/3} = \bar{f}^{-2/3}_\mathrm{b} \cdot 4 \pi G \rho^{1/3}_\mathrm{s} r^2_\mathrm{s}\cdot L(\tilde{r}), 
\end{equation}
where
\begin{equation}
    K(\tilde{r}) \equiv \int^\infty_{\tilde{r}} x^{-3} (1+x)^{-2}\left[ \mathrm{ln}(1+x) - \frac{x}{1+x}\right] \mathrm{d}x,
\end{equation}
and
\begin{equation}
    L(\tilde{r}) \equiv \tilde{r}^{5/3}(1+\tilde{r})^{10/3}K(\tilde{r}).
\end{equation}
Over the radial range of interest here, $\tilde{r} \in (10^{-2}, 10^2\,)$, $L(\tilde{r})$ is represented to better than 10\% by the simple power law, $0.255\,\tilde{r}^{1.28}$. Thus, $S(r)$ can be taken to be a power law with this index.\footnote{This behaviour agrees with that found by \citet{Taylor&Navarro2001} for the pseudo-phase-space density of isotropic NFW dark matter haloes (which is $\propto S^{-1.5}$) after allowing for the different radial ranges fitted in the two studies.} 

Using the halo mass definition adopted throughout this paper, eq. \ref{eq:sr} can be converted into 
\begin{equation}
\begin{split}
    S(r) = & \bar{f}^{-2/3}_\mathrm{b} \cdot (12 \pi)^{1/3} c^{-1} \left[ \mathrm{ln}(1+c) - \frac{c}{1+c}\right]^{-1/3} \cdot \\ 
    & G\rho^{-1/3}_\mathrm{200}M^{2/3}_\mathrm{200} \cdot L(\tilde{r}),
\end{split}
\end{equation}
where $\rho_{200} \equiv 200 \Omega_\mathrm{m} \rho_\mathrm{crit,~0}(1+z)^3$ is the mean matter density within the halo virial radius $r_{200}$, the critical density of the Universe at present is $\rho_\mathrm{crit,~0} \equiv 3{H^2_0}/8\pi G$, and $c \equiv  r_{200}/r_{\mathrm s}$ is the halo concentration. 

The initial entropy $S_\mathrm{i}$ is given by
\begin{equation}
S_\mathrm{i} \equiv p(z_\mathrm{i})\rho(z_\mathrm{i})^{-5/3} = 0.6 \left(\bar{f}_\mathrm{b}\,\Omega_\mathrm{m}\,\rho_\mathrm{crit,\,0}\right)^{-2/3} c^2_\mathrm{s,\,0}\,, 
\end{equation}
where $c_\mathrm{s,\,0}=\sqrt{5k_\mathrm{B}T_0/3\mu m_\mathrm{p}}$,  in which $k_\mathrm{B}$ is the Boltzmann constant, $T_0$ denotes $T_\mathrm{i}(1+z_\mathrm{i})^{-2}$ with $T_\mathrm{i}$ representing the initial gas temperature (i.e. $T_\mathrm{i}=245~\mathrm{K}$ for the NR, NR-H, Ln, Fn, and Hn runs, and $T_{\mathrm i}=10^{7}~\mathrm{K}$ for the Lf, Ff, and Hf runs), and $\mu$ is the mean molecular weight of a physical gas particle in units of the proton mass, $m_\mathrm{p}$. For neutral gas composed of 76\% hydrogen and 24\% helium, $\mu=1.2195$ and $c_\mathrm{s,\,0}=0.013 (T_\mathrm{i}/245\,\mathrm{K})^{1/2}~ \mathrm{km}\cdot \mathrm{s}^{-1}$. 

Normalising the entropic function profile by $S_{\mathrm i}$, we thus obtain 
\begin{equation}
\label{eq:Sscale}
    \frac{S(r)}{S_\mathrm{i}} = \frac{c^{0.28}}{\left[\ln(1+c)-c/(1+c)\right]^{1/3}}\left(\frac{r}{r_\mathrm{200}}\right)^{1.28}\left(\frac{M_\mathrm{200}}{M_*}\right)^{2/3},
\end{equation}
where $M_*$ is defined by
\begin{equation}
\label{eq:m*}
    M_* = 24.06 \left(G H_0\right)^{-1} \Omega_\mathrm{m}^{-1/2} c^3_\mathrm{s,\,0}  (1+z)^{3/2}.
\end{equation}

Equation \ref{eq:Sscale} is only valid at radii greater than a core radius defined by $S(r_{\mathrm c}) = S_{\mathrm i}$. At smaller radii we assume the gas to be adiabatic with $S=S_\mathrm{i}$, so that $p(r) = S_{\mathrm i} \rho_{\mathrm g}(r)^{5/3}$. Integrating eq. \ref{eq:eql} in this case gives
\begin{equation}
    \phi(r_\mathrm{c}) - \phi(r) = \frac{5}{2} \left[ \frac{p(r)}{\rho_\mathrm{g}(r)} - \frac{p(r_\mathrm{c})}{\rho_\mathrm{g}(r_\mathrm{c})} \right].
\end{equation}
Using eq.~\ref{eq:potential} then gives the gas density profile at $r\leq r_{\mathrm c}$,
\begin{equation}
\label{eq:rhogri}
\begin{split}
    \rho_\mathrm{g}(r) & = \bar{f}_\mathrm{b} \cdot \rho_\mathrm{s} \tilde{r}^{-1}_\mathrm{c} (1+\tilde{r}_\mathrm{c})^{-2} \cdot \\
    & \left[1 + \frac{2}{5} \frac{\tilde{r}^{-1} \ln(1 + \tilde{r}) - \tilde{r}_\mathrm{c}^{-1} \ln(1 + \tilde{r}_\mathrm{c})}{\tilde{r}_\mathrm{c} (1+\tilde{r}_\mathrm{c})^{2} K(\tilde{r}_\mathrm{c})}\right]^{3/2}, 
\end{split}
\end{equation} 
where $\tilde{r}_\mathrm{c} \equiv r_\mathrm{c}/r_\mathrm{s}$ is the nondimensionalized core radius. 

Integrating the gas density profile (eq. \ref{eq:rhogri} at $r\leq r_{\mathrm c}$ and eq. \ref{eq:rhogro} at $r\geq r_{\mathrm c}$) over $0<r<r_{200}$, and dividing by $\bar{f}_\mathrm{b}\;\!M_{200}$, we obtain the factor $F$ by which the halo baryonic fraction $f_\mathrm{b,\,halo}$ is reduced relative to the cosmic baryon fraction $\bar{f}_\mathrm{b}$, 
\begin{equation}
\label{eq:fbar}
\begin{split}
    F \left(\frac{M_{200}}{M_*}, c\right)  & \equiv f_\mathrm{b,\,halo}/\bar{f}_\mathrm{b}\\
    & = \ 1 - \left[ \mathrm{ln}(1+c) - \frac{c}{1+c}\right]^{-1} \cdot \\ 
    & \ \ \ \int^{\mathrm{min}(\tilde{r}_\mathrm{c},\,c)}_0 \!\! \big(\tilde{r}^{-1} (1 + \tilde{r})^{-2} - \! \tilde{r}^{-1}_\mathrm{c}(1 + \tilde{r}_\mathrm{c})^{-2} \cdot \\
    & \ \ \ \left[1 + \frac{2}{5} \frac{\tilde{r}^{-1} \ln(1 + \tilde{r}) - \tilde{r}_\mathrm{c}^{-1} \ln(1 + \tilde{r}_\mathrm{c})}{\tilde{r}_\mathrm{c} (1+\tilde{r}_\mathrm{c})^{2} K(\tilde{r}_\mathrm{c})}\right]^{3/2} \big) \, \tilde{r}^2\mathrm{d}\tilde{r}, 
\end{split}
\end{equation}
where $\tilde{r}_\mathrm{c}$ is given by $S(r_\mathrm{c})/S_\mathrm{i}=1$, hence,
\begin{equation}
    \tilde{r}_\mathrm{c} = \left[ c^{-1}\left(\ln(1+c) - c/(1+c)\right)^{-1/3} \left( \frac{M_\mathrm{200}}{M_*} \right)^{2/3} \right]^{-1/1.28}, 
\end{equation}
and to keep consistency, $K(\tilde{r}_\mathrm{c})$ is approximated by  
\begin{equation}
\begin{split}
    K(\tilde{r}_\mathrm{c}) & = \tilde{r}_\mathrm{c}^{-5/3}(1+\tilde{r}_\mathrm{c})^{-10/3}L(\tilde{r}_\mathrm{c}) \\
    & \approx 0.255 \tilde{r}_\mathrm{c}^{1.28}\cdot \tilde{r}_\mathrm{c}^{-5/3}(1+\tilde{r}_\mathrm{c})^{-10/3}.
\end{split}
\end{equation}

Setting $F=0.5$ in eq. \ref{eq:fbar}, we can solve for the characteristic mass $M_{1/2}$, at which haloes have a baryonic fraction which is half of the cosmic mean. Over the range $c \in [0.1, 100]$ we find a good numerical fit to the result to be 
\begin{equation}
\label{eq:mhalf}
\begin{split}
    \log_{10}\frac{M_\mathrm{1/2}}{M_*} =  & -0.491 + 0.489\,\log_{10}c -0.0988 (\log_{10}c)^2 \\ & + 0.0495 (\log_{10}c)^3 
    -0.00589 (\log_{10}c)^4.
\end{split}
\end{equation}
Since $M_*=327 \,(1+z)^{3/2} \, (T_\mathrm{i}/245\,\mathrm{K})^{3/2}\, \mathrm{M_\odot}$ (for NR, NR-H runs), the dependence of baryonic fraction on halo mass in eq.~\ref{eq:fbar}, and the characteristic mass\footnote{We provide a code at \url{https://github.com/haonan-zheng/model\_fbar\_halo} for readers to estimate the baryonic fraction and the characteristic mass $M_{1/2}$, particularly the mass scale to consider effects of the thermal pressure arising from the initial gas entropy with any halo concentration, redshift, and cosmology.} of eq.~\ref{eq:mhalf} are easily compared with the simulations discussed in the main text. Note that $M_{1/2}$ increases with $c$ because the nondimensionalized core radius increases with halo concentration. Between $c=5$ and $35$, $M_\mathrm{1/2}/M_*$ increases from 0.658 to 1.507. In Table \ref{tab:tableA1}, we compare our model with the simulations in the main text, and find it to be reasonably accurate although always somewhat low; in the worst case (for NR/NR-H at $z=9.27$) the prediction is about 50\% lower than the characteristic mass estimated from the simulation. 

\begin{table}
 \caption{Comparison of our model for the characteristic mass $M_{1/2}$ at which haloes have half the cosmic baryonic fraction to non-radiative simulations with different initial gas temperatures $T_\mathrm{i}$ at different times. $M_\mathrm{1/2,\,simulation}$ is interpolated/extrapolated from the runs with the highest resolution (i.e. NR-H and Hf); note that for the NR, Ln, Fn, and Hn runs, the drop in baryonic fraction is spurious. $M_\mathrm{1/2,\,model}$ is predicted using eq. \ref{eq:mhalf} with halo concentration $c=35, 10, 5$ for NR-H\protect\footnotemark[6], and with $c=5$ for Hn and Hf (i.e. $M_\mathrm{1/2,\,model}/M_*=1.507$, $0.878$ and $0.658$ respectively): for the NR and NR-H runs, $M_*=327(1+z)^{3/2}$; for the other runs, we update $M_*$  with eq. \ref{eq:m*} according to the appropriate cosmological parameters, initial temperature, and mean gas molecular weight. Our model provides reasonably accurate predictions. }
 \label{tab:tableA1} 
 
\begin{tabular}{cccccc}
  \hline
  \shortstack{Name of\\simulations} & \shortstack{\vspace{0.125cm} $z$} & \shortstack{\vspace{0.06cm} $T_\mathrm{i} [\mathrm{K}]$} & \shortstack{\vspace{0.150cm}$c$} &\shortstack{$M_{\mathrm{1/2,\,simulation}}$ \\ $[\mathrm{M_{\odot}}]$} & \shortstack{$M_{\mathrm{1/2,\,model}}$\\$[\mathrm{M_{\odot}}]$} \\
  \hline

   & $0.00$ & $245$ & $35$ & -- & $4.92 \times 10^2$ \\
   & $3.07$ & $245$ & $10$ & $3.78 \times 10^3$ & $2.35 \times 10^3$ \\
  \shortstack{\vspace{0.03cm}NR, NR-H} & $5.72$ & $245$ & $5$ & $6.96 \times 10^3$ & $3.74 \times 10^3$ \\
   & $9.27$ & $245$ & $5$ & $1.33 \times 10^4$ & $7.07 \times 10^3$ \\
  Ln, Fn, Hn & $1.97$ & $245$ & $5$ & -- & $1.08 \times 10^3$ \\ 
  Lf, Ff, Hf & $1.97$ & $10^7$ & $5$ & $3.54 \times 10^{10}$  & $2.65 \times 10^{10}$ \\ 

  \hline
 \end{tabular}
 
\end{table}

\footnotetext[6]{\label{fn}These values are estimated from the further zoomed NR-H run and dark-matter-only runs in \citet{Wang2020}.}

%%%%%%%%%%%%%%%%%%%%%%%%%%%%%%%

% Don't change these lines
\bsp	% typesetting comment
\label{lastpage}
\end{document}